\documentclass[a4paper]{jpconf}

\usepackage{graphicx}
\usepackage{epstopdf}
\usepackage{dcolumn}
\usepackage{amsmath}
\usepackage{amssymb}

\newcommand{\ba}{\begin{eqnarray}}
\newcommand{\ea}{\end{eqnarray}}
\newcommand{\bmath}{\begin{mathletters}}
\newcommand{\emath}{\end{mathletters}}
\newcommand{\ban}{\begin{eqnarray*}}
\newcommand{\ean}{\end{eqnarray*}}

\newcommand{\bsub}{\begin{subequations}}
\newcommand{\esub}{\end{subequations}}

\def\ket#1{|#1\rangle}
	
\def\bsu3{\overline{{\rm SU(3)}}}
\def\bso6{\overline{{\rm SO(6)}}}
\def\bPi2{\overline{\Pi}^{(2)}}
\def\blam{\bar{\lambda}}
\def\bmu{\bar{\mu}}
\def\bK{\bar{K}}
\def\b0{\beta_0}
\def\beq{\beta_{\rm eq}}
\def\g0{\gamma_0}
\def\gaeq{\gamma_{\rm eq}}
\def\bs{\beta_*}
\def\bss{\beta_{**}}

\begin{document}

\title{Partial dynamical symmetries and shape coexistence in nuclei}

\author{A. Leviatan and N. Gavrielov}

\address{Racah Institute of Physics, The Hebrew University, 
Jerusalem 91904, Israel}

\ead{ami@phys.huji.ac.il, noam.gavrielov@mail.huji.ac.il}


\begin{abstract}
We present a symmetry-based approach for shape coexistence in nuclei, 
founded on the concept of partial dynamical symmetry (PDS). 
The latter corresponds 
to a situation when only selected states (or bands of states) 
of the coexisting configurations preserve the symmetry while other 
states are mixed. We construct explicitly critical-point Hamiltonians 
with two or three PDSs of the type U(5), SU(3), $\bsu3$ and SO(6), 
appropriate to double or triple coexistence of spherical, prolate, 
oblate and $\gamma$-soft deformed shapes, respectively. 
In each case, we analyze the topology of the energy surface 
with multiple minima and corresponding normal modes. Characteristic features 
and symmetry attributes of the quantum spectra and wave functions are 
discussed. Analytic expressions for quadrupole moments and $E2$ rates 
involving the remaining solvable states are derived and isomeric states 
are identified by means of selection rules.
\end{abstract}
\vspace{-0.8cm}

\section{Introduction}
\label{intro}

The presence in the same nuclei, at similar low energies, of two or more 
sets of states which have distinct properties that can be interpreted in 
terms of different shapes, is a ubiquitous 
phenomena across the nuclear chart~\cite{Heyde11,Focus16}. 
The increased availability of rare isotope beams and advancement in 
high-resolution spectroscopy, open new capabilities to investigate such 
phenomena in nuclei far from stability~\cite{Jenkins14}.
Notable empirical examples include the coexistence of prolate and 
oblate shapes in the neutron-deficient Kr~\cite{Clement07}, 
Se~\cite{Ljun08} and Hg~\cite{Bree14} isotopes and in the 
neutron-rich Se isotopes~\cite{Chen17}, 
the coexistence of spherical and deformed shapes in neutron-rich Sr 
isotopes~\cite{Clement16,Park16}, $^{96}$Zr~\cite{kremer16} 
and near $^{78}$Ni~\cite{gottardo16,yang16}, and the triple coexistence 
of spherical, prolate and oblate shapes in $^{186}$Pb~\cite{Andreyev00}. 
A~detailed microscopic interpretation of nuclear shape-coexistence
is a formidable task. In a shell model description of nuclei 
near shell-closure, it is attributed to the occurrence of 
multi-particle multi-hole intruder excitations across shell gaps. 
For medium-heavy nuclei, this necessitates drastic truncations of 
large model spaces, {\it e.g.}, by Monte Carlo 
sampling~\cite{MCSM99,Tsunuda14}
or by a bosonic approximation of nucleon pairs~[16-25]. 
In a mean-field approach, based on energy density functionals, 
the coexisting shapes are associated with different minima 
of an energy surface calculated self-consistently. 
A detailed comparison with spectroscopic observables 
requires beyond mean-field methods, 
including restoration of broken symmetries and configuration mixing of 
angular-momentum and particle-number projected 
states~\cite{Bender13,vretenar16}. 
Such extensions present a major computational effort and often 
require simplifying assumptions such as axial symmetry 
and/or a mapping to collective model 
Hamiltonians~[23-28].

A recent global mean-field calculation of nuclear shape isomers 
identified experimentally accessible regions of nuclei 
with multiple minima in their potential-energy 
surface~\cite{moller09,moller12}.  
Such heavy-mass nuclei awaiting exploration, 
are beyond the reach of realistic large-scale shell model calculations. 
With that in mind, we present a simple 
alternative to describe shape coexistence in medium-heavy nuclei, 
away from shell-closure, in the 
framework of the interacting boson model (IBM)~\cite{ibm}. 
The proposed approach emphasizes the role of remaining underlying 
symmetries which provide physical insight and make the problem tractable.
The feasibility of such a symmetry-based approach 
gains support from the previously proposed~\cite{Iac00,Iac01} 
and empirically confirmed~\cite{Cas00,Cas01} analytic descriptions of 
critical-point nuclei.

\section{Dynamical symmetries and nuclear shapes}
\label{ds-shapes}

The IBM has been widely used to describe low-lying quadrupole 
collective states in nuclei in terms of $N$ 
monopole ($s^\dag$) and quadrupole ($d^\dag$) bosons,
representing valence nucleon pairs. 
The model has U(6) as a spectrum generating algebra, where 
the Hamiltonian is expanded in terms of its generators, 
$\{s^{\dag}s,\,s^{\dag}d_{m},\, d^{\dag}_{m}s,\,d^{\dag}_{m}d_{m '}\}$,
and consists of Hermitian, rotational-scalar interactions 
which conserve the total number of $s$- and $d$- bosons, 
${\textstyle \hat N = \hat{n}_s + \hat{n}_d = 
s^{\dagger}s + \sum_{m}d^{\dagger}_{m}d_{m}}$. 
The solvable limits of the model correspond to dynamical symmetries 
associated with chains of nested sub-algebras of U(6), 
terminating in the invariant SO(3) algebra. A dynamical 
symmetry~(DS) occurs when the Hamiltonian is expressed in terms 
of the Casimir operators of a given chain,
\ba
{\rm U(6)\supset G_1\supset G_2\supset \ldots \supset SO(3)} &&\;\;\quad
\ket{N,\, \lambda_1,\,\lambda_2,\,\ldots,\,L} ~.\quad 
\label{u6-ds}
\ea
In such a case, all states are solvable and classified by 
quantum numbers, $\ket{N,\, \lambda_1,\,\lambda_2,\,\ldots,\,L}$, 
which are the labels of irreducible representations (irreps) of the 
algebras in the chain. Analytic expressions are available 
for energies and other observables and definite selection rules for 
transition processes. 
The DS chains with leading sub-algebras ${\rm G_1}$: 
U(5), SU(3), ${\rm\overline{SU(3)}}$ and SO(6), 
correspond to known paradigms of nuclear collective 
structure: spherical vibrator, prolate-, oblate- and $\gamma$-soft deformed 
rotors, respectively. 

A geometric visualization of the IBM is obtained by an energy surface
\ba
E_{N}(\beta,\gamma) &=& 
\langle \beta,\gamma; N\vert \hat{H} \vert \beta,\gamma ; N\rangle ~, 
\label{enesurf}
\ea
defined by the expectation value of the Hamiltonian in the coherent 
(intrinsic) state~\cite{gino80,diep80},
\bsub
\ba
\vert\beta,\gamma ; N \rangle &=&
(N!)^{-1/2}(b^{\dagger}_{c})^N\,\vert 0\,\rangle ~,
\\[1mm]
b^{\dagger}_{c} &=& (1+\beta^2)^{-1/2}[\beta\cos\gamma 
d^{\dagger}_{0} + \beta\sin{\gamma} 
( d^{\dagger}_{2} + d^{\dagger}_{-2})/\sqrt{2} + s^{\dagger}] ~. 
\ea
\label{int-state}
\esub
Here $(\beta,\gamma)$ are
quadrupole shape parameters whose values, $(\beta_{\rm eq},\gamma_{\rm eq})$, 
at the global minimum of $E_{N}(\beta,\gamma)$ define the equilibrium 
shape for a given Hamiltonian. 
The shape can be spherical $(\beta \!=\!0)$ or 
deformed $(\beta \!>\!0)$ with $\gamma \!=\!0$ (prolate), 
$\gamma \!=\!\pi/3$ (oblate), 
$0 \!<\! \gamma \!<\! \pi/3$ (axially asymmetric) or $\gamma$-independent.
The equilibrium deformations associated with the 
DS limits conform with their geometric interpretation 
and are given by 
$\beta_{\rm eq}\!=\!0$ for U(5), 
$(\beta_{\rm eq} \!=\!\sqrt{2},\gamma_{\rm eq}\!=\!0)$ for SU(3), 
$(\beta_{\rm eq} \!=\!\sqrt{2},\gamma_{\rm eq}\!=\!\pi/3)$ for $\bsu3$, 
and $(\beta_{\rm eq}\!=\!1,\gamma_{\rm eq}\,\,{\rm arbitrary})$ for SO(6). 
The DS Hamiltonians support a single minimum in their 
energy surface, hence serve as benchmarks for the dynamics of a single 
quadrupole shape (spherical, axially-deformed and $\gamma$-unstable deformed).

\section{Partial dynamical symmetries and multiple nuclear shapes}
\label{pds-shapes}

A dynamical symmetry (DS) is characterized by {\it complete} solvability 
and good quantum numbers for {\it all} states. Partial dynamical 
symmetry (PDS)~\cite{Leviatan11,Alhassid92,Leviatan96} 
is a generalization of the latter concept, and corresponds 
to a particular symmetry breaking for which 
only {\it some} of the states retain solvability 
and/or have good quantum numbers. 
Such generalized forms of symmetries are manifested in nuclear structure, 
where extensive tests provide empirical evidence for their 
relevance to a broad range of nuclei~[38, 40-57]. 
In addition to nuclear spectroscopy, Hamiltonians with PDS 
have been used in the study of 
quantum phase transitions~\cite{Leviatan07,Macek14,LevDek16} and of 
systems with mixed regular and chaotic dynamics~\cite{WAL93,LW93}. 
In the present work, we show that this novel symmetry notion 
can play a vital role in formulating algebraic benchmarks for 
the dynamics of multiple quadrupole shapes. 

Coexistence of different shapes involve 
several states (or bands of states) with 
distinct properties, reflecting the 
nature of their dissimilar dynamics. 
The relevant Hamiltonians, by necessity, contain competing terms 
with incompatible (non-commuting) symmetries, 
hence exact dynamical symmetries are broken. 
In the IBM, the required symmetry breaking is achieved by including in the 
Hamiltonian terms associated with different DS chains, {\it e.g.}, 
by mixing the Casimir operators of the 
leading sub-algebra in each chain~\cite{diep80}. 
This mixing and the resulting quantum phase transitions have been studied 
extensively in the IBM 
framework~\cite{IacZamCas98,IacZam04,lev05,Leviatan06,CejJolCas10,Iac11}. 
In general, under such circumstances, solvability is lost, 
there are no remaining non-trivial conserved quantum numbers and all 
eigenstates are expected to be mixed. Shape coexistence near shell closure 
was considered within the IBM with configuration mixing, 
by using different Hamiltonians for the normal 
and intruder configurations and a number-non-conserving mixing 
term~\cite{DuvBar81,DuvBar82,SambMoln82,Foisson03,Frank04,Morales08}. 
In the present work, we adapt a different strategy. We construct 
a single number-conserving Hamiltonian with PDS,  
which retains the virtues of the relevant dynamical symmetries, but only 
for selected sets of states associated with each shape. 
We focus on the dynamics in the vicinity of the critical 
point where the corresponding multiple minima in the energy surface 
are near-degenerate and the structure changes most rapidly. 
The construction relies on an intrinsic-collective resolution of the 
Hamiltonian~\cite{kirlev85,Leviatan87,levkir90}, 
a procedure used formerly in the study of 
first-order quantum phase transitions~\cite{Leviatan06}. 

The above indicated resolution amounts to separating the complete 
Hamiltonian $\hat{H}'=\hat{H}+\hat{H}_c$ 
into an intrinsic part ($\hat{H}$), which determines the energy surface, 
and a collective part ($\hat{H}_c$), 
which is composed of kinetic rotational terms. 
For a given shape, specified by the equilibrium deformations 
$(\beq,\gaeq)$, the intrinsic Hamiltonian is required 
to annihilate the equilibrium intrinsic state, Eq.~(\ref{int-state}),
\ba
\hat{H}\ket{\beq,\gaeq;N} = 0 ~.
\label{Ncond}
\ea
Since the Hamiltonian is rotational-invariant, 
this condition is equivalent to the requirement 
that $\hat{H}$ annihilates the states of good 
angular momentum~$L$ projected from $\ket{\beq,\gaeq;N}$
\ba
\hat{H}\ket{\beq,\gaeq;N,x,L} = 0 ~.
\label{Nxl}
\ea
Here $x$ denotes additional quantum numbers needed to characterize the 
states and, for simplicity, we have omitted the irrep label $M$ of 
${\rm SO(2)}\subset{\rm SO(3)}$.
Symmetry considerations enter when $(\beq,\gaeq)$ coincide 
with the equilibrium deformations of the DS chains, 
mentioned in Section~\ref{ds-shapes}.
In this case, the equilibrium intrinsic state, $\ket{\beq,\gaeq;N}$, 
becomes a lowest (or highest) weight state in a particular irrep, 
$\lambda_1=\Lambda_0$, of the leading 
sub-algebra $G_1$ in the chain of Eq.~(\ref{u6-ds}). 
The projected states,  
$\ket{\beq,\gaeq;N,\lambda_1\!=\!\Lambda_0,\lambda_2,\ldots,L}$, 
are now specified by the quantum numbers of the algebras in the chain 
and the intrinsic Hamiltonian $\hat{H}$ satisfies
\ba
\hat{H}\ket{\beq,\gaeq;N,\lambda_1\!=\!\Lambda_0,\lambda_2,\ldots,L} = 0 ~.
\label{Hvanish}
\ea 
The set of zero-energy eigenstates in Eq.~(\ref{Hvanish}) 
are basis states of the particular $G_1$-irrep, $\lambda_1=\Lambda_0$, 
and have good $G_1$ symmetry. 
For a positive-definite $\hat{H}$, they span the ground band of 
the equilibrium shape. $\hat{H}$ itself, however, need not be invariant 
under $G_1$ and, therefore, has partial-$G_1$ symmetry. 
Identifying the collective part with the Casimir operators of 
the remaining sub-algebras of $G_1$ in the chain~(\ref{u6-ds}), 
the degeneracy of the above set of states 
is lifted, and they remain solvable eigenstates of the complete Hamiltonian. 
The latter, by definition, has $G_1$-PDS. 
According to the PDS algorithms~\cite{Alhassid92,GarciaRamos09}, 
the construction of number-conserving Hamiltonians obeying the condition  
of Eq~(\ref{Hvanish}), is facilitated by writing them in 
normal-order form, $\hat{H} = 
\sum_{\alpha,\beta}u_{\alpha\beta}\hat{T}^{\dag}_{\alpha}\hat{T}_{\beta}$, 
in terms of $n$-particle creation and annihilation operators satisfying 
\ba
\hat{T}_{\alpha}
\ket{\beq,\gaeq;N,\lambda_1\!=\!\Lambda_0,\lambda_2,\ldots,L} = 0 ~.
\label{Talpha}
\ea

A large number of purely 
bosonic~\cite{Leviatan11,Alhassid92,Leviatan96,LevSin99,
LevIsa02,LevGino00,GarciaRamos09,Leviatan13}, 
purely fermionic~\cite{Escher00,Escher02} and 
bose-fermi~\cite{Pds-BF15} Hamiltonians with PDS have been constructed in 
this manner. 
With a few exceptions~\cite{Leviatan07,Macek14,LevDek16}, 
they all involved a single PDS. 
We now wish to extend the above procedure to encompass a construction 
of Hamiltonians with several distinct PDSs, relevant to coexistence 
of multiple shapes. For that purpose, consider two different shapes 
specified by equilibrium deformations 
($\beta_1,\gamma_1$) and ($\beta_2,\gamma_2$) whose dynamics is 
described, respectively, by the following DS chains
\bsub
\ba
{\rm U(6)\supset G_1\supset G_2\supset \ldots \supset SO(3)} &&\;\;\quad
\ket{N,\, \lambda_1,\,\lambda_2,\,\ldots,\,L} ~,\quad 
\label{ds-G1}\\
{\rm U(6)\supset G'_1\supset G'_2\supset \ldots \supset SO(3)} &&\;\;\quad
\ket{N,\, \sigma_1,\,\sigma_2,\,\ldots,\,L} ~,\quad 
\label{ds-G1prime}
\ea
\esub
with different leading sub-algebras ($G_1\neq G'_1$) and associated bases.
At the critical point, the corresponding minima representing the two shapes 
and the respective ground bands are degenerate. Accordingly, 
we require the intrinsic critical-point Hamiltonian to satisfy 
simultaneously the following two conditions
\bsub
\ba
\hat{H}\ket{\beta_1,\gamma_1;N,\lambda_1\!=\Lambda_0,\lambda_2,\ldots,L} 
&=& 0 ~,
\label{basis1}\\
\hat{H}\ket{\beta_2,\gamma_2;N,\sigma_1=\Sigma_0,\sigma_2,\ldots,L} 
&=&0 ~.
\label{basis2}
\ea
\label{bases12}
\esub
The states of Eq.~(\ref{basis1}) reside in the $\lambda_1=\Lambda_0$ irrep 
of $G_1$, are classified according to the DS-chain (\ref{ds-G1}), hence 
have good $G_1$ symmetry. Similarly,  
the states of Eq.~(\ref{basis2}) reside in the $\sigma_1=\Sigma_0$ irrep 
of $G'_1$, are classified according to the DS-chain (\ref{ds-G1prime}), 
hence have good $G'_1$ symmetry. Although $G_1$ and $G'_2$ are incompatible, 
both sets are eigenstates of the same Hamiltonian. When the latter 
is positive definite, the two sets span the ground bands of the 
$(\beta_1,\gamma_1)$ and $(\beta_2,\gamma_2)$ shapes, respectively.
In general, $\hat{H}$ itself is not necessarily 
invariant under $G_1$ nor under $G_2$ and, therefore, its other eigenstates 
can be mixed under both $G_1$ and $G'_1$. 
Identifying the collective part of the Hamiltonian with the Casimir 
operator of SO(3) (as well as with the Casimir operators of additional 
algebras which are common to both chains), the two sets of states 
remain (non-degenerate) eigenstates of the complete Hamiltonian which 
then has both $G_1$-PDS and $G'_1$-PDS. 
The case of triple (or multiple) 
shape coexistence, associated with three (or more) incompatible DS-chains is 
treated in a similar fashion. 
In the following Sections, we apply this procedure 
to a variety of coexisting shapes, examine the spectral properties 
of the derived PDS Hamiltonians, and highlight their potential to serve 
as benchmarks for describing multiple shapes in nuclei. 

\section{Spherical and axially-deformed shape coexistence: U(5)-SU(3) PDS}
\label{SPshapes}

A particular type of shape coexistence present in nuclei, 
involves spherical and axially-deformed shapes. 
The relevant DS chains for such configurations are~\cite{ibm},
\bsub
\ba
&&
{\rm U(6)\supset U(5)\supset SO(5)\supset SO(3)} \;\;\quad
\ket{N,\, n_d,\,\tau,\,n_{\Delta},\,L} ~,\quad 
\label{u5SP}
\\
&&
{\rm U(6)\supset SU(3)\supset SO(3)} \;\,\,\qquad\quad\quad
\ket{N,\, (\lambda,\mu),\,K,\, L} ~.\quad 
\label{su3SP}
\ea
\label{u5su3}
\esub
The U(5)-DS limit of Eq.~(\ref{u5SP}) 
is appropriate to the dynamics of a spherical shape. 
For a given U(6) irrep~$N$, the allowed U(5) and SO(5) irreps 
are $n_d\!=\!0,1,2,\ldots, N$ and  $\tau\!=\!n_d,\,n_d\!-\!2,\dots 0$ 
or~$1$, respectively. The values of $L$ contained in a given $\tau$-irrep 
follow the ${\rm SO(5)\supset SO(3)}$ reduction rules~\cite{ibm} 
and $n_{\Delta}$ is a multiplicity label.
The basis states, $\ket{N,\, n_d,\,\tau,\,n_{\Delta},\,L}$, 
are eigenstates of the Casimir operators 
$\hat{C}_{1}[{\rm U(5)}]\!=\!\hat{n}_d$, 
$\hat{C}_{2}[{\rm U(5)}]\!=\!\!\hat{n}_d(\hat{n}_d+4)$, 
$\hat{C}_{2}[{\rm SO(5)}]\!=\sum_{\ell=1,3}U^{(\ell)}\cdot U^{(\ell)}$ and 
$\hat{C}_{2}[{\rm SO(3)}]=L^{(1)}\cdot L^{(1)}$,  
with eigenvalues $n_d$, $n_d(n_d+4)$, $\tau(\tau+3)$ and $L(L+1)$, 
respectively. 
Here $\hat{C}_{k}[{\rm G}]$ denotes the Casimir operator of ${\rm G}$ 
of order $k$, 
$\hat{n}_d\!=\!\sqrt{5}\,U^{(0)}$, $L^{(1)}\!=\!\sqrt{10}\,U^{(1)}$, 
with $U^{(\ell)}\!=\!(d^{\dag}\tilde{d})^{(\ell)}$, 
$\tilde{d}_{m} \!=\! (-1)^{m}d_{-m}$ and standard notation 
of angular momentum coupling is used. 
The U(5)-DS Hamiltonian involves a linear combination of these 
Casimir operators. 
The spectrum resembles that of an anharmonic spherical vibrator, 
describing quadrupole excitations of a spherical shape. 
The splitting of states in a given U(5) $n_d$-multiplet 
is governed by the SO(5) and SO(3) terms. 
The lowest U(5) multiplets involve the ground state 
with quantum numbers $(n_d\!=\!0,\,\tau\!=\!0,\, L\!=\!0)$ 
and excited states with quantum numbers 
$(n_d=\!1\!,\,\tau\!=\!1,\, L\!=\!2)$, 
$(n_d\!=\!2,\,\tau\!=\!0,\,L\!=\!0;\,\tau\!=\!2,\,L\!=\!2,4)$ 
and  $(n_d\!=\!3,\,\tau\!=\!3,\,L\!=\!6,4,3,0;\,\tau\!=\!1,\,L\!=\!2)$. 

The SU(3)-DS limit of Eq.~(\ref{su3SP}) is appropriate
to the dynamics of a prolate-deformed shape. 
For a given $N$, the allowed SU(3) irreps are 
$(\lambda,\mu)\!=\!(2N \!-\! 4k \!-\! 6m,2k)$ 
with $k,m$, non-negative integers. 
The values of $L$ contained in a given $(\lambda,\mu)$-irrep 
follow the ${\rm SU(3)\supset SO(3)}$ reduction rules~\cite{ibm} 
and the multiplicity label $K$ corresponds geometrically to the
projection of the angular momentum on the symmetry axis. 
The basis states are eigenstates of the Casimir operator 
$\hat{C}_{2}[SU(3)] \!=\! 2Q^{(2)}\cdot Q^{(2)} \!+\! 
{\textstyle\frac{3}{4}}L^{(1)}\cdot L^{(1)}$ 
with eigenvalues $\lambda(\lambda+3)+\mu(\mu+3)+\lambda\mu$.
The generators of SU(3) are the angular momentum operators $L^{(1)}$ 
defined above, and the quadrupole operators
\ba
Q^{(2)} = d^{\dagger}s + s^{\dagger}\tilde{d} 
-\frac{1}{2}\sqrt{7} (d^{\dagger}\tilde{d})^{(2)} ~.
\label{Qsu3}
\ea
The SU(3)-DS Hamiltonian involves a linear combination of 
$\hat{C}_{2}[SU(3)]$ and $\hat{C}_{2}[SO(3)]$, and its 
spectrum resembles that of an axially-deformed 
rotovibrator composed of SU(3) $(\lambda,\mu)$-multiplets forming 
rotational bands with $L(L+1)$-splitting. 
The lowest irrep $(2N,0)$ contains the ground band $g(K\!=\!0)$ 
of a prolate deformed nucleus. 
The first excited irrep $(2N\!-\!4,2)$ contains 
both the $\beta(K\!=\!0)$ and $\gamma(K\!=\!2)$ bands. 

In discussing properties of the SU(3)-DS spectrum, it is convenient 
to subtract from $\hat{C}_{2}[{\rm SU(3)}]$ the ground-state energy, and 
consider the following positive-definite term
\ba
\hat{\theta}_2 \equiv
-\hat C_{2}[{\rm SU(3)}] + 2\hat{N} (2\hat{N} +3) =
P^{\dagger}_{0}P_{0} + P^{\dagger}_{2}\cdot \tilde{P}_{2} ~, 
\label{theta2}
\ea 
where $\tilde{P}_{2m} = (-)^{m}P_{2,-m}$. The SU(3) basis states, 
$\ket{N,\, (\lambda,\mu),\,K,\, L}$, 
are eigenstates of $\hat{\theta}_2$ with eigenvalues 
$(2N-\lambda)(2N+\lambda+3)- \mu(\lambda+\mu+3)$ 
and the ground 
band with $(\lambda,\mu)=(2N,0)$ occurs at zero energy.
The two-boson pair operators 
\bsub
\ba
P^{\dagger}_{0} &=& d^{\dagger}\cdot d^{\dagger} - 2(s^{\dagger})^2 ~,
\label{P0}
\\
P^{\dagger}_{2m} &=& 2d^{\dagger}_{m}s^{\dagger} + 
\sqrt{7}\, (d^{\dagger}\,d^{\dagger})^{(2)}_{m} ~,
\label{P2}
\ea
\esub
are $(0,2)$ tensors with respect to SU(3) and satisfy 
\ba
P_{0}\,\ket{N,\, (\lambda,\mu)\!=\!(2N,0),\,K\!=\!0,\, L} &=& 0 ~,
\nonumber\\
P_{2m}\,\ket{N,\, (\lambda,\mu)\!=\!(2N,0),\,K\!=\!0,\, L} &=& 0 ~.
\label{P0P2}
\ea
These operators correspond to 
$\hat{T}_{\alpha}$ of Eq.~(\ref{Talpha}) and, as shown below, they play 
a central role in the construction of Hamiltonians with SU(3)-PDS.
\begin{figure}[t]
  \centering
\includegraphics[width=16cm]{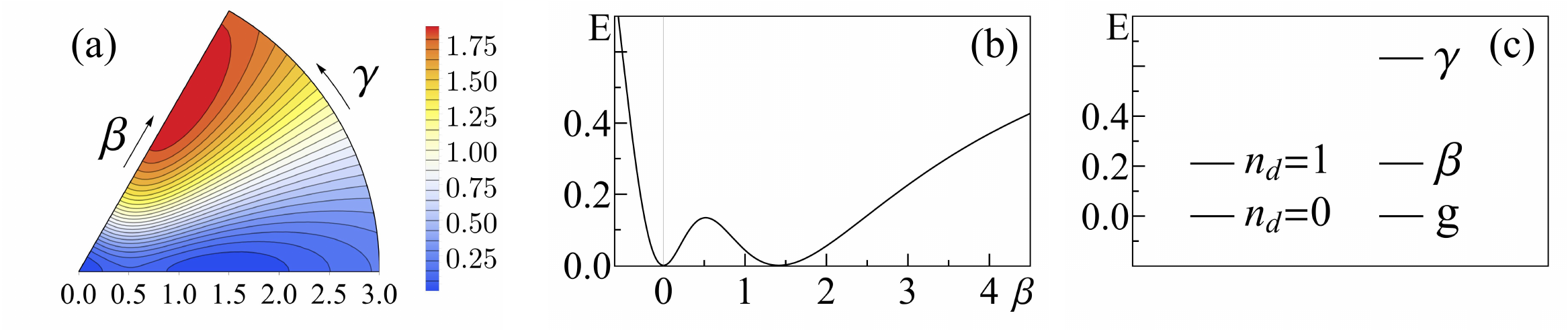}
  \caption{
Spherical-prolate (S-P) shape coexistence.
(a)~Contour plots of the energy surface~(\ref{Ebgu5su3}),   
(b)~$\gamma\!=\!0$ sections, and
(c)~bandhead spectrum, for the Hamiltonian $\hat{H}'$~(\ref{HprimeSP}) 
with parameters $h_2\!=\!1,\,\rho=0$ and $N\!=\!20$. 
\label{fig1SP}}
\end{figure}

Considering the case of coexisting spherical and prolate-deformed shapes, 
following the procedure outlined in Eq.~(\ref{bases12}), 
the intrinsic part of the 
critical-point Hamiltonian is required to satisfy
\bsub
\ba
\hat{H}\ket{N,\, (\lambda,\mu)\!=\!(2N,0),\,K\!=\!0,\, L} &=& 0 
\qquad\qquad L=0,2,4,\ldots,2N
\label{2N0}
\\
\hat{H}\ket{N,\, n_d=0,\,\tau=0,\,L=0} &=& 0 ~.
\label{nd0}
\ea
\label{u5su3vanish}
\esub 
Equivalently, $\hat{H}$ annihilates both the intrinsic state of 
Eq.~(\ref{int-state}) with $(\beta=\sqrt{2},\gamma=0)$, 
which is the lowest weight vector in the SU(3) irrep 
$(\lambda,\mu)=(2N,0)$, and the intrinsic state with $\beta=0$, 
which is the single basis state in the U(5) irrep $n_d=0$. 
The resulting intrinsic critical-point 
Hamiltonian is found to be~\cite{Leviatan07},
\ba
\hat{H} = h_{2}\,P^{\dagger}_{2}\cdot \tilde{P}_{2} ~, 
\label{Hu5su3}
\ea
where $P^{\dagger}_{2m}$ is given in Eq.~(\ref{P2}). 
The corresponding energy surface, 
$E_{N}(\beta,\gamma) = N(N-1)\tilde{E}(\beta,\gamma)$, 
is given by
\ba
\tilde{E}(\beta,\gamma) &=& 
2h_2\beta^2\,
(\,\beta^2 - 2\sqrt{2}\beta\cos 3\gamma +2\,)(1+\beta^2)^{-2} ~.
\label{Ebgu5su3}
\ea
The surface is linear in $\Gamma\!=\!\cos 3\gamma$, at most quartic 
in $\beta$, and can be transcribed as 
$\tilde{E}(\beta,\gamma) = 
(1+\beta^2)^{-2}(a\beta^2-b\beta^3\Gamma + c\beta^4)$, 
with $b^2=4ac$ and $a=4h_2,\,b=4\sqrt{2}h_2,\,c=2h_2$. 
It is the most general form 
of a surface accommodating 
degenerate spherical and axially-deformed extrema, 
for an Hamiltonian with one- and two-body terms. 
For $h_2> 0$, $\hat{H}$ is positive definite and 
$\tilde{E}(\beta,\gamma)$ has two degenerate global minima, 
$\beta=0$ and $(\beta=\sqrt{2},\gamma=0)$, at $\tilde{E}=0$. 
Additional extremal points include 
(i)~a saddle point: 
$[\beta\!=\!\bs \!=\! (\sqrt{3}-1)/\sqrt{2},\gamma\!=\!0]$, 
which supports a barrier of height 
$\tilde{E}_{\rm bar} \!=\!\frac{1}{2}h_2(\sqrt{3}-1)^2=0.268h_2$, 
and (ii)~a~local maximum: 
$[\beta=\bss = (\sqrt{3}+1)/\sqrt{2},\gamma=\pi/3]$ 
[or equivalently $(\beta \!=\! -\bss,\gamma\!=\!0)]$, at 
$\tilde{E}_{\rm max} = \frac{1}{2}h_2(\sqrt{3}+1)^2$.
Figs.~1(a) and 1(b) show the energy surface contour,  
$\tilde{E}(\beta,\gamma)$, and section, $\tilde{E}(\beta,\gamma\!=\!0)$, 
of $\hat{H}$, respectively. The barrier separating the two minima satisfies 
$\tilde{E}_{\rm bar}=0.13\tilde{E}_{\rm lim}$, 
where 
$\tilde{E}_{\rm lim}\!=\!
\tilde{E}(\beta\!\rightarrow\!\infty,\gamma)\!=\!2h_2$. 
It is significantly higher than typical barriers 
of order $0.001\tilde{E}_{\rm lim}$, obtained in 
standard Hamiltonians mixing the U(5) and SU(3) Casimir 
operators~\cite{lev05}. 
The normal modes of the Hamiltonian~(\ref{Hu5su3}) correspond to 
small oscillations about the respective minima of its energy surface. 
For large $N$, the deformed normal modes involve 
one-dimensional $\beta$ vibration and two-dimensional 
$\gamma$ vibrations about the prolate-deformed global minimum, with 
frequencies
\bsub
\ba
\epsilon_{\beta} &=& 4h_{2}N ~,
\label{betmodes}\\
\epsilon_{\gamma} &=& 12h_{2}N ~.
\label{gammodes}
\ea
\label{betgammodes}
\esub
The spherical normal modes involve five-dimensional quadrupole vibrations 
about the spherical global minimum, with frequency
\ba
\epsilon = 4h_{2}N ~.
\label{sphermodes}
\ea
The bandhead spectrum associated with these modes, 
is shown in Fig.~(1c).
Interestingly, the spherical and $\beta$ modes have the same energy 
and are considerably lower than the $\gamma$ mode, 
$\epsilon=\epsilon_{\beta} = \epsilon_{\gamma}/3$. This is consistent with 
the observed enhanced density of low-lying $0^{+}$ states, 
signaling the transitional region of 
such a first-order quantum phase transition~\cite{Cejnar00,Meyer06}.

By construction, the members of the prolate-deformed ground-band 
$g(K=0)$, Eq.~(\ref{2N0}), have good SU(3) quantum numbers 
$(\lambda,\mu)=(2N,0)$, and the spherical ground state, Eq.~(\ref{nd0}), 
has good U(5) quantum numbers $(n_d\!=\!\tau\!=\!L\!=\!0)$. 
The Hamiltonian $\hat{H}$ 
of Eq.~(\ref{Hu5su3}) has additional 
solvable SU(3) basis states with 
$(\lambda,\mu)\!=\!(2N-4k,2k)K\!=\!2k$, which span the deformed 
$\gamma^k(K\!=\!2k)$ bands, and an additional solvable U(5) basis state 
with $n_d\!=\!\tau\!=\!L\!=\!3$. 
Altogether, although $\hat{H}$ 
is neither SU(3)-invariant nor U(5)-invariant, it has a subset of 
solvable states with good SU(3) symmetry 
\bsub
\ba
&\vert N,(2N,0)K=0,L\rangle\qquad\; &  E\!=\!0  
\qquad\qquad\qquad\quad\; 
L=0,2,4,\ldots, 2N ~,
\label{gK0}
\\
&\ket{N,(2N\!-\!4k,2k)K\!=\!2k,L} &   E  \!=\!  
h_{2}\, 6k (2N \!-\! 2k\!+\!1 ) 
\quad L\!=\!K,K\!+\!1,..., (2N\!-\!2k) ~,\qquad
\label{gammaK}
\ea
\label{su3solv}
\esub
and, simultaneously, a subset of solvable states with good U(5) symmetry
\bsub
\ba
& \ket{N, n_d=\tau=L=0} \quad\;\; & E = 0 ~,
\label{L0}\\
& \ket{N, n_d=\tau=L=3}\quad\;\; & E = 6(2N -1) ~.
\label{L3}
\ea
\label{u5solv}
\esub 
The spherical $L=0$ state, Eq.~(\ref{L0}), 
is degenerate with the prolate-deformed ground band, Eq.~(\ref{gK0}), 
and the spherical $L=3$ state, Eq.~(\ref{L3}), 
is degenerate with the $\gamma$ band, Eq.~(\ref{gammaK}) with $k=1$. 
Identifying the collective part with $\hat{C}_2[{\rm SO(3)}]$, 
we arrive at the following complete Hamiltonian 
\ba
\hat{H}' &=& h_{2}\,P^{\dagger}_{2}\cdot \tilde{P}_{2} 
+ \rho\,\hat{C}_2[\rm SO(3)] ~.
\label{HprimeSP}
\ea 
The added rotational term generates an exact $L(L+1)$ splitting without 
affecting the wave functions. In particular, the solvable subsets of 
eigenstates, Eq.~(\ref{su3solv})-(\ref{u5solv}), remain intact. 
Other eigenstates, as shown below, can mix strongly 
with respect to both SU(3) and U(5). 

The symmetry structure of the Hamiltonian eigenstates 
can be inferred from the probability distributions, 
${\textstyle P^{(N,L)}_{n_d} = \sum_{\tau,n_{\Delta}}
\vert C_{n_d,\tau,n_{\Delta}}^{(N,L)}\vert^2}$ and 
${\textstyle P^{(N,L)}_{(\lambda,\mu)} = \sum_{K}
\vert C_{(\lambda,\mu),K}^{(N,L)}\vert^2}$, obtained from 
their expansion coefficients in the U(5) and SU(3) 
bases~(\ref{u5su3}), respectively. 
In general, the low lying spectrum of $\hat{H}'$~(\ref{HprimeSP}) 
exhibits two distinct type of states, spherical and deformed. 
Spherical type of states show
a narrow $n_d$-distribution, with a characteristic dominance of 
a single $n_d$ component. 
Fig.~2 shows the U(5) $n_d$-decomposition of such states, 
selected on the
basis of having the largest components with $n_d=0,1,2,3$,
within the given $L$ spectra. States with different $L$ values
are arranged into panels labeled by `$n_d$' to conform with the
structure of the $n_d$-multiplets of a spherical vibrator. 
The lowest spherical $L=0^{+}_2$ state is seen to be a pure $n_d=0$ state
which is the solvable U(5) state of Eq.~(\ref{L0}). 
The $L=2^{+}_2$ state has a pronounced $n_d=1$ component 
($\sim$~80\%), whose origin 
can be traced to the relation
\ba
&&
\hat{H}\ket{N,n_d\!=\!\tau\!=\!1,L\!=\!2} =
\nonumber\\ 
&&\qquad\qquad
{\textstyle
h_2\,4(N-1)\left [\ket{N,n_d\!=\!\tau\!=\!1,L\!=\!2} 
+ \sqrt{\frac{7}{2(N-1)}}\,\ket{N,n_d=\tau=L=2}\right ]} ~.\qquad
\label{Hnd1SP}
\ea
As seen, the U(5)-basis state $\ket{N,n_d\!=\!\tau\!=\!1,L\!=\!2}$ 
approaches the status of an eigenstate 
for large $N$, with corrections of order 
$1/\sqrt{N}$. The states ($L=0^{+}_4,\,2^{+}_6,\,4^{+}_3$) 
in the third panel of Fig.~2, have a less pronounced 
($\sim$~50\%)
single $n_d=2$ component. The higher-energy states
in the panel `$n_d=3$' are significantly fragmented, 
with a notable exception of $L=3^{+}_2$, which is the 
solvable U(5) basis state of Eq.~(\ref{L3}).
\begin{figure}[t]
  \centering
\includegraphics[width=15cm]{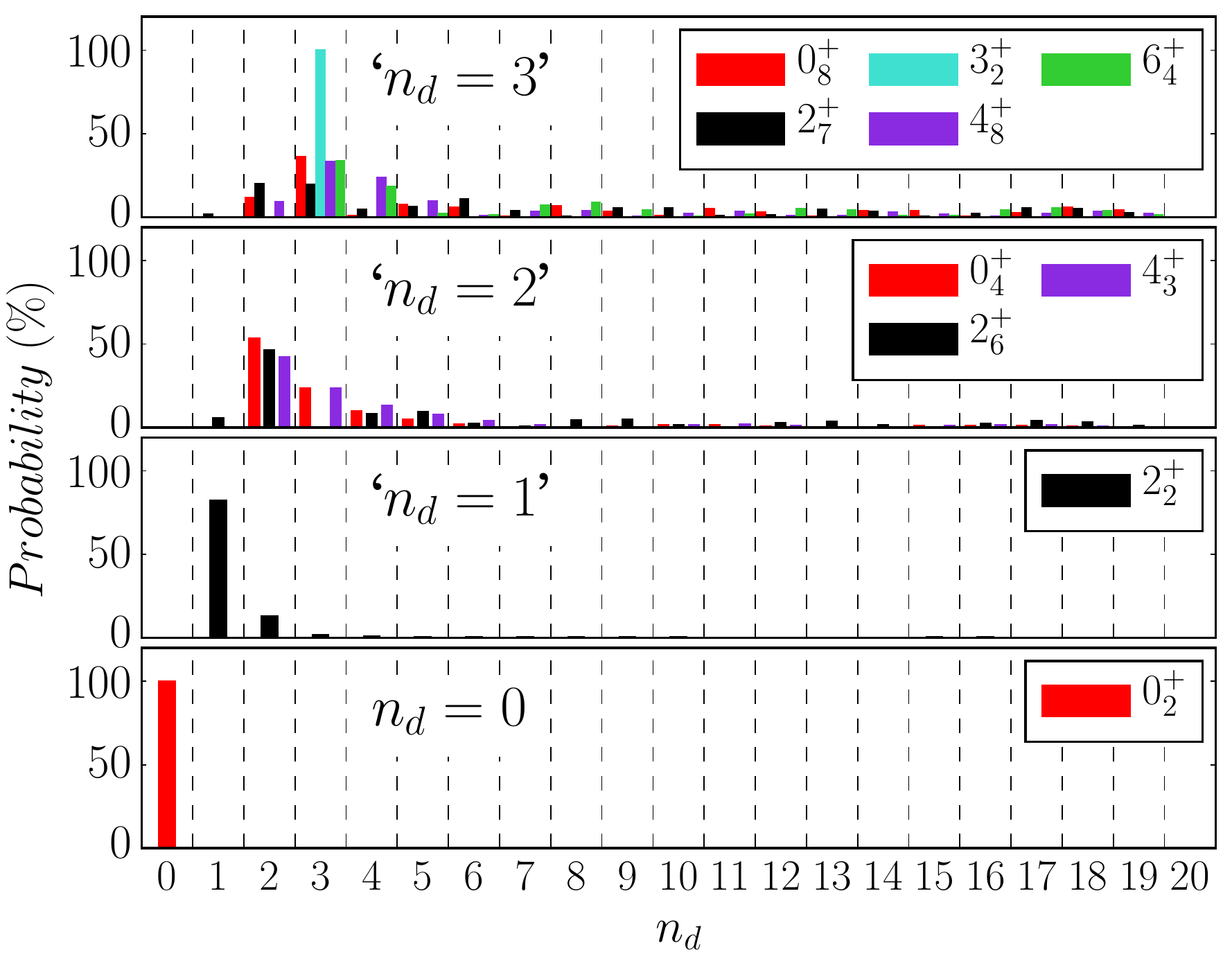}
\caption{
\label{fig2SP}
U(5) $n_d$-decomposition for spherical states, 
eigenstates of 
the Hamiltonian $\hat{H}'$~(\ref{HprimeSP}) 
with parameters as in Fig.~1, resulting 
in spherical-prolate shape coexistence.}
\end{figure}
\begin{figure}[t]
  \centering
\includegraphics[width=15cm]{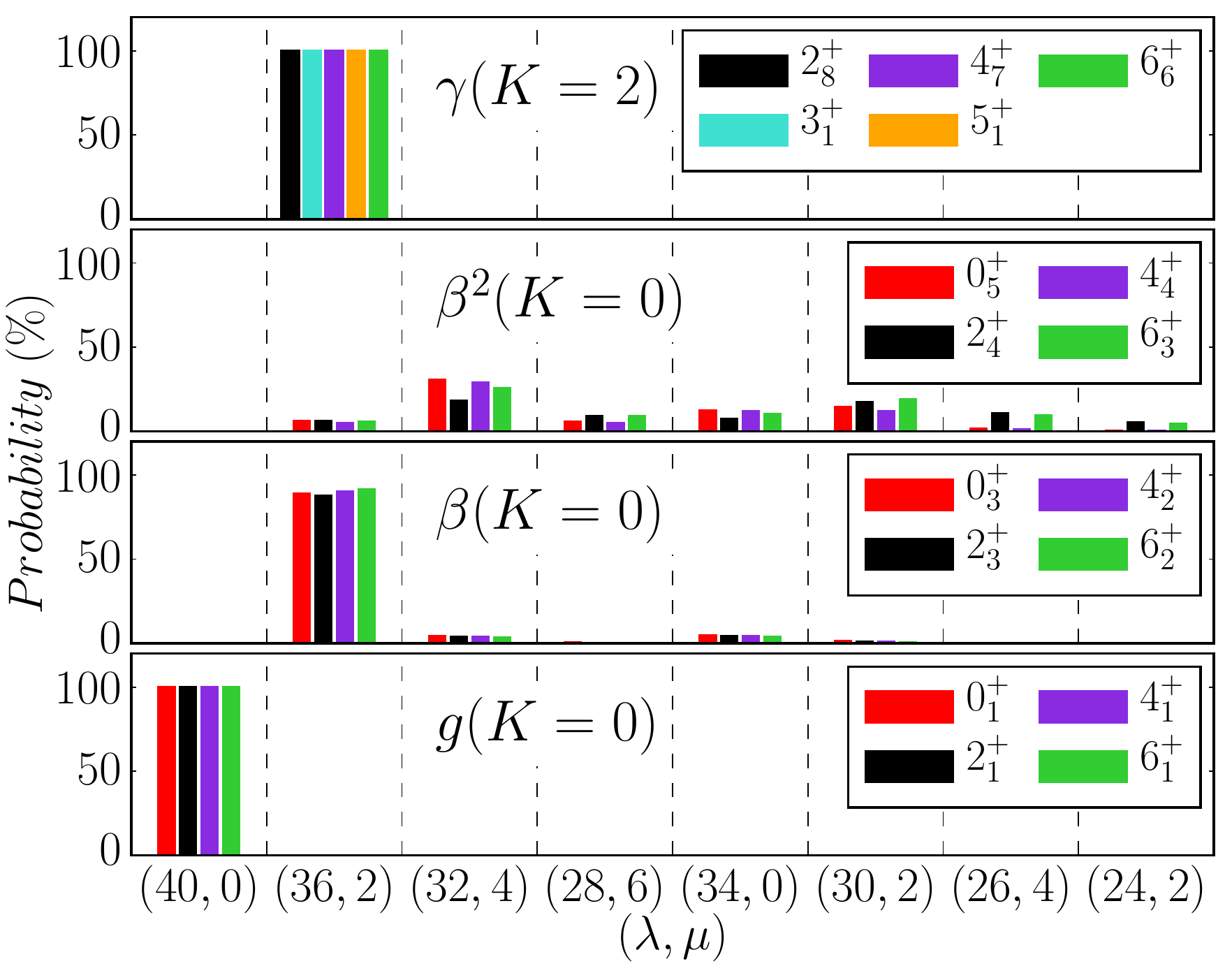}
\caption{
\label{fig3SP}
SU(3) $(\lambda,\mu)$-decomposition
for members of the prolate-deformed
$g,\,\beta,\,\beta^2$ and $\gamma$ bands, 
eigenstates of the Hamiltonian $\hat{H}'$~(\ref{HprimeSP}) 
with parameters as in Fig.~1. 
Shown are probabilities larger than 4\%.} 
\end{figure}
 
The deformed type of states have a different character. They 
exhibit a broad $n_d$-distribution, as 
seen clearly in the following expansion 
of the SU(3) ground band wave functions in the U(5) basis
\bsub
\ba
&&\ket{N,(2N,0)K=0,L} = 
\sum_{n_d,\tau,n_{\Delta}}
\frac{1}{2}
\left [1 + (-1)^{n_d-\tau}\right ]\, 
\xi_{n_d,\tau,n_{\Delta}}^{(N,L)}
\ket{N,n_d,\tau,n_{\Delta},L} ~,\qquad\;\;\\
&&
{\textstyle
\xi_{n_d,\tau,n_{\Delta}}^{(N,L)} =
\left[\frac{N!(2N-L)!!(2N+L+1)!!}{3^{N}(2N)!
(N-n_d)!(n_d-\tau)!!(n_d +\tau +3)!!}
\right ]^{1/2}
(\sqrt{2})^{n_d}\,f_{\tau,n_{\Delta}}^{(L)}} ~.\qquad
\ea
\label{g1u5expan}
\esub
Explicit expressions of $f_{\tau,n_{\Delta}}^{(L)}$ for $L\!=\!0,2,4$ 
are documented in~\cite{Leviatan11}. 
The U(5) $n_d$-probability inferred from Eq.~(\ref{g1u5expan}), 
shows that the contribution of 
each individual $n_d$-component is exponentially small for large $N$. 
Fig.~3 shows the SU(3) $(\lambda,\mu)$-distribution
for such deformed-type of states, 
members of the $g(K\!=\!0),\,\beta(K\!=\!0),\,\beta^2(K\!=\!0)$ 
and $\gamma(K\!=\!2)$ bands. 
The ground $g(K\!=\!0)$ and $\gamma(K\!=\!2)$
bands are pure with $(\lambda,\mu) = (2N,0)$ and $(2N-4,2)$
SU(3) character, respectively. These are the solvable bands of
Eq.~(\ref{su3solv}) with good SU(3) quantum numbers. 
The non-solvable $K$-bands, {\it e.g.} the $\beta(K\!=\!0)$ 
and $\beta^2(K\!=\!0)$ in Fig.~3, show considerable SU(3) mixing, and the 
mixing is coherent, {\it i.e.}, similar for different $L$-states
in the same band. The above analysis demonstrates that some eigenstates of 
the critical-point Hamiltonian~(\ref{HprimeSP}) have good U(5) symmetry 
(either exactly or to a good approximation for large~$N$), 
some eigenstates have good SU(3) symmetry, 
and all other states are mixed with respect to both U(5) and SU(3). 
This defines U(5)-PDS coexisting with SU(3)-PDS.
These persisting competing symmetries affect the dynamics at the 
critical point, which has a mixed regular 
and chaotic character~\cite{Macek14}.

Since the wave functions for the solvable states, 
Eqs.~(\ref{su3solv})-(\ref{u5solv}), are known, one has at hand closed form 
expressions for electromagnetic moments and rates. 
Taking the $E2$ operator to be proportional to the 
SU(3) quadrupole operator of Eq.~(\ref{Qsu3}) 
with an effective charge $e_B$, $T(E2) = e_B\,Q^{(2)}$, 
the $B(E2)$ values for intraband ($g\to g$) 
transitions between states of the ground band~(\ref{gK0}) 
and quadrupole moments are given by the known SU(3)-DS expressions~\cite{ibm} 
\bsub
\ba
Q_L &=& 
{\textstyle
-e_B\sqrt{\frac{16\pi}{40}}\frac{L}{2L+3}(4N+3)} ~,
\label{QLsu3}\\[1mm]
B(E2; g,\, L+2\to g,\,L) &=& 
{\textstyle
e_{B}^2\,\frac{3(L+1)(L+2)}{4(2L+3)(2L+5)}
(2N-L)(2N+L+3)} ~.
\qquad\qquad
\label{be2Su3}
\ea
\label{QLbe2Su3}
\esub
Similarly, the quadrupole moment of the 
solvable spherical $L\!=\!3$ state of Eq.~(\ref{L3}), 
obeys the U(5)-DS expression
\ba
Q_{L=3} 
&=& 
{\textstyle
-e_B3\sqrt{\frac{16\pi}{40}}} ~.
\label{QL3u5}
\ea
The spherical states, Eq.~(\ref{u5solv}), are not connected by 
$E2$ transitions to states of the ground band~(\ref{gK0}), 
since the latter exhaust the $(2N,0)$ irrep of SU(3) and 
$Q^{(2)}$, as a generator, cannot connect different 
$(\lambda,\mu)$-irreps of SU(3). As will be discussed in Section~6, 
weak spherical$\,\to\,$deformed $E2$ transitions 
persist also for a more general $E2$ operator, 
obtained by adding to $T(E2)$ the term $d^{\dag}s+s^{\dag}\tilde{d}$. 
The latter, however, can connect by $E2$ transitions the ground 
with excited $\beta$ and $\gamma$ bands. Since both the $g(K=0)$ and 
$\gamma(K=2)$ bands are 
solvable with good SU(3) symmetry, Eq.~(\ref{su3solv}), 
analytic expressions for $\gamma\to g$ $E2$ rates are 
available~\cite{Leviatan96,Isa83}. 
There are also no $E0$ transitions involving the 
spherical states~(\ref{u5solv}), since the $E0$ operator 
$T(E0)\propto \hat{n}_d$, is diagonal in $n_d$.

The above discussion has focused on the dynamics in the vicinity of the 
critical point where the spherical and deformed minima are near degenerate. 
The evolution of structure away from the critical point can be studied 
by incorporating additional terms into $\hat{H}'$ (\ref{HprimeSP}). 
Adding an $\epsilon\hat{n}_d$ term, 
will leave the solvable spherical states~(\ref{u5solv}) 
unchanged, but will shift the deformed ground band to higher energy 
of order $2\epsilon N/3$. 
Similarly, adding a small $\alpha\hat{\theta}_2$ term, Eq.~(\ref{theta2}), 
will leave the solvable SU(3) bands unchanged but will shift the spherical 
ground state ($n_d\!=\!L\!=\!0$) to higher energy of order $4\alpha N^2$. 
The selection rules discussed above, ensure that the $L\!=\!0$ state of 
the excited configuration will have 
significantly retarded $E2$ and $E0$ decays to states of the lower 
configuration, hence will have the attributes of an isomer state.

\section{Prolate-oblate shape coexistence: 
SU(3)-$\mathbf{\overline{SU(3)}}$ PDS}
\label{POshapes}

Shape coexistence in nuclei can involve two deformed shapes, 
{\it e.g.}, prolate and oblate. The relevant DS limits 
for the latter configurations are~\cite{ibm},
\bsub
\ba
{\rm U(6)\supset SU(3)\supset SO(3)} &&\;\;\quad
\ket{N,\, (\lambda,\mu),\,K,\, L} ~,\quad 
\label{SU3}
\\
{\rm U(6)\supset \bsu3\supset SO(3)} &&\;\;\quad
\ket{N,\, (\blam,\bmu),\,\bar{K},\, L} ~.\quad
\label{SU3bar}
\ea
\label{su3chains}
\esub
The SU(3)-DS chain~(\ref{SU3}), appropriate to a prolate-shape, 
was discussed in Section~\ref{SPshapes}. 
The $\bsu3$-DS chain~(\ref{SU3bar}), appropriate to 
an oblate-shape, has similar properties but now 
the allowed $\bsu3$ irreps are 
$(\blam,\bmu)\!=\!(2k,2N\!-\!4k\!-\!6m)$ 
with $k,m$, non-negative integers, and the multiplicity label is 
denoted by $\bK$. 
The basis states are eigenstates of the Casimir operator 
$\hat{C}_{2}[\bsu3] \!=\! 2\bar{Q}^{(2)}\cdot \bar{Q}^{(2)} \!+\! 
{\textstyle\frac{3}{4}}L^{(1)}\cdot L^{(1)}$, 
with eigenvalues $\blam(\blam+3)+\bmu(\bmu+3)+\blam\bmu$.
Here $\bar{Q}^{(2)}$ are the quadrupole operators given by
\ba
\bar{Q}^{(2)} = d^{\dagger}s + s^{\dagger}\tilde{d} 
+\frac{1}{2}\sqrt{7} (d^{\dagger}\tilde{d})^{(2)} ~,
\label{Qsu3bar}
\ea
and $L^{(1)}$ are the angular momentum operators. 
The generators of SU(3) and $\bsu3$, $Q^{(2)}$~(\ref{Qsu3}) 
and $\bar{Q}^{(2)}$~(\ref{Qsu3bar}), 
and corresponding basis states, 
$\ket{N,(\lambda,\mu),K,L}$ and $\ket{N,(\blam,\bmu),\bar{K},L}$, 
are related by a change of phase $(s^{\dag},s)\rightarrow (-s^{\dag},-s)$, 
induced by the operator 
${\cal R}_s=\exp(i\pi\hat{n}_s)$, with $\hat{n}_s=s^{\dag}s$. 
As previously mentioned, in the SU(3)-DS, 
the prolate ground band  $g(K\!=\!0)$ has 
$(2N,0)$ character and the $\beta(K\!=\!0)$ and $\gamma(K\!=\!2)$ 
bands have $(2N\!-\!4,2)$. 
In the $\bsu3$-DS, the oblate ground band $g(\bK\!=\!0)$ has 
$(0,2N)$ character and the excited $\beta(\bK\!=\!0)$ and 
$\gamma(\bK\!=\!2)$ bands
have $(2,2N\!-\!4)$. 
Henceforth, we denote such prolate and oblate bands by 
$(g_1,\beta_1,\gamma_1)$ and ($g_2,\beta_2,\gamma_2$), respectively. 
Since ${\cal R}_sQ^{(2)}{\cal R}_s^{-1} \!=\! -\bar{Q}^{(2)}$, 
the SU(3) and $\bsu3$ DS spectra are identical and 
the quadrupole moments of corresponding states differ in sign.

The phase transition between prolate and oblate shapes has been 
previously studied by varying a control parameter 
in the IBM Hamiltonian~\cite{jolie01,jolie03}. 
The latter, however, consisted of one- and two-body terms 
hence could not accommodate simultaneously two deformed minima. 
For that reason, in the present approach, 
we consider an Hamiltonian with cubic terms which retains the virtues 
of SU(3) and $\bsu3$ DSs for the prolate and oblate ground bands.
Following the procedure of Eq.~(\ref{bases12}), 
the intrinsic part of such critical-point Hamiltonian is required to~satisfy
\bsub
\ba
\hat{H}\ket{N,\, (\lambda,\mu)=(2N,0),\,K=0,\, L} &=& 0 ~,
\label{2N0p}
\\
\hat{H}\ket{N,\, (\blam,\bmu)=(0,2N),\,\bar{K}=0,\, L} &=& 0 ~.
\label{02N}
\ea
\label{HvanishPO}
\esub 
Equivalently, $\hat{H}$ annihilates the intrinsic states of 
Eq.~(\ref{int-state}), with $(\beta\!=\!\sqrt{2},\gamma\!=\!0)$ and 
$(\beta\!=\!-\sqrt{2},\gamma\!=\!0)$, which are the lowest- and 
highest-weight vectors in the irreps $(2N,0)$ and $(0,2N)$ 
of SU(3) and $\bsu3$, respectively. 
The resulting Hamiltonian is found to be~\cite{LevDek16},
\ba
\hat{H} = 
h_0\,P^{\dag}_0\hat{n}_sP_0 + h_2\,P^{\dag}_0\hat{n}_dP_0 
+\eta_3\,G^{\dag}_3\cdot\tilde{G}_3 ~,
\label{HintPO}
\ea
where $P^{\dag}_{0}$ is given in Eq.~(\ref{P0}), 
$G^{\dag}_{3,\mu} = \sqrt{7}[(d^{\dag} d^{\dag})^{(2)}d^{\dag}]^{(3)}_{\mu}$, 
$\tilde{G}_{3,\mu} = (-1)^{\mu}G_{3,-\mu}$. 
The corresponding energy surface, 
$E_{N}(\beta,\gamma) = N(N-1)(N-2)\tilde{E}(\beta,\gamma)$, 
is given by
\ba
\tilde{E}(\beta,\gamma) = 
\left\{(\beta^2-2)^2
\left [h_0 + h_2\beta^2\right ] 
+\eta_3 \beta^6\sin^2(3\gamma)\right \}(1+\beta^2)^{-3} ~.
\label{surface}
\ea
\begin{figure}[t]
  \centering
\includegraphics[width=16cm]{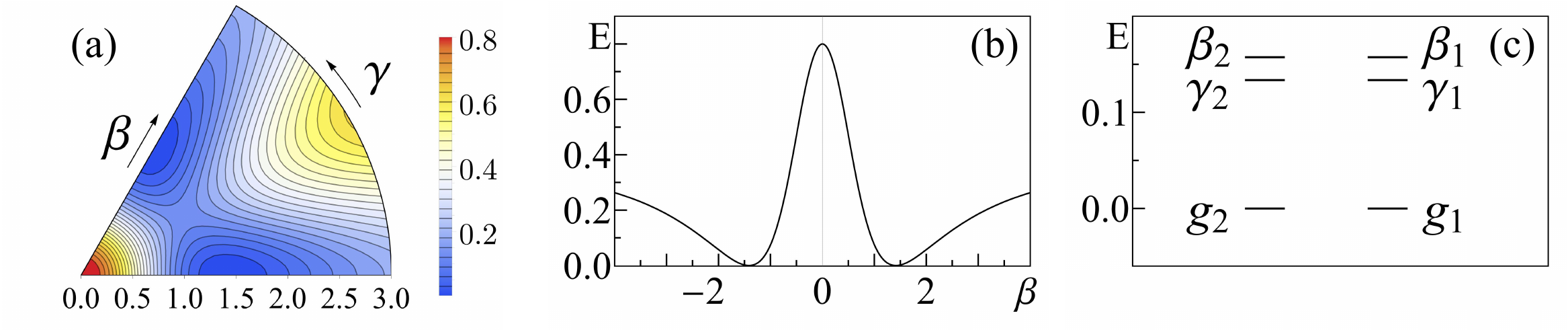}
  \caption{
Prolate-oblate (P-O) shape coexistence.
(a)~Contour plots of the energy surface~(\ref{surface}),   
(b)~$\gamma\!=\!0$ sections, and
(c)~bandhead spectrum, for the Hamiltonian $\hat{H}'$~(\ref{HprimePO})
with parameters $h_0\!=\!0.2,\,h_2\!=\!0.4,\,\eta_3\!=\!0.567,
\,\alpha\!=\!0.018,\,\rho=0$ and $N\!=\!20$. 
\label{fig4PO}}
\end{figure}
The surface is an even function of $\beta$ and 
$\Gamma = \cos 3\gamma$, 
and can be transcribed as 
$\tilde{E}(\beta,\gamma) = z_0 + 
(1+\beta^2)^{-3}[A\beta^6+ B\beta^6\Gamma^2 + D\beta^4+ F\beta^2]$, 
with $A \!=\! -4h_0 \!+\!h_2 \!+\! \eta_3,\, B \!=\! -\eta_3, \,
D \!=\! -(11h_0 \!+\! 4h_2), \; F \!=\! 4(h_2\!-\!4h_0),\,z_0 \!=\! 4h_0$. 
It is the most general form 
of a surface accommodating 
degenerate prolate and oblate extrema
with equal $\beta$-deformations, 
for an Hamiltonian with cubic terms~\cite{isacker81,levshao90}. 
For $h_0,h_2,\eta_3\geq 0$, 
$\hat{H}$ is positive definite and 
$\tilde{E}(\beta,\gamma)$ has two degenerate global minima, 
$(\beta=\sqrt{2},\gamma=0)$ and $(\beta=\sqrt{2},\gamma=\pi/3)$ 
[or equivalently $(\beta=-\sqrt{2},\gamma=0)$], at $\tilde{E}=0$.
$\beta=0$ is always an extremum, which is a local minimum (maximum) for 
$F>0$ ($F<0$), at $\tilde{E}=4h_0$.
Additional extremal points include 
(i)~a saddle point: $[\bs^2 = \frac{2(4h_0-h_2)}{h_0 - 7h_2}, 
\gamma=0,\pi/3]$, at $\tilde{E}=\frac{4(h_0+2h_2)^3}{81(h_0-h_2)^2}$.
(ii)~A~local maximum and a saddle point: $[\bss^2,\gamma=\pi/6]$, 
at $\tilde{E}= \frac{1}{3}(1+\bss^2)^{-2}\bss^2[D\bss^2+2F] +z_0$, 
where $\bss^2$ is a solution of 
$(D-3A)\bss^4 + 2(F-D)\bss^2 -F = 0$. 
The saddle points, when they exist, support 
a barrier separating the various minima, as seen in Fig.~4. 
For large $N$, the normal modes involve 
$\beta$ and $\gamma$ vibrations about the 
respective deformed minima, with frequencies
\bsub
\ba
&&
\epsilon_{\beta 1}=\epsilon_{\beta 2} 
= \frac{8}{3}(h_0+ 2h_2)N^2 ~,
\\
&&\epsilon_{\gamma 1}=\epsilon_{\gamma 2} = 4\eta_3N^2 ~.
\ea
\label{d-modes}
\esub
Figs.~(4a), (4b) and (4c) show 
$\tilde{E}(\beta,\gamma)$, $\tilde{E}(\beta,\gamma\!=\!0)$ 
and the bandhead spectrum, 
respectively, with parameters ensuring prolate-oblate (P-O) global 
minima and a local maximum at $\beta=0$.
\begin{figure}[t]
  \centering
\includegraphics[width=16cm]{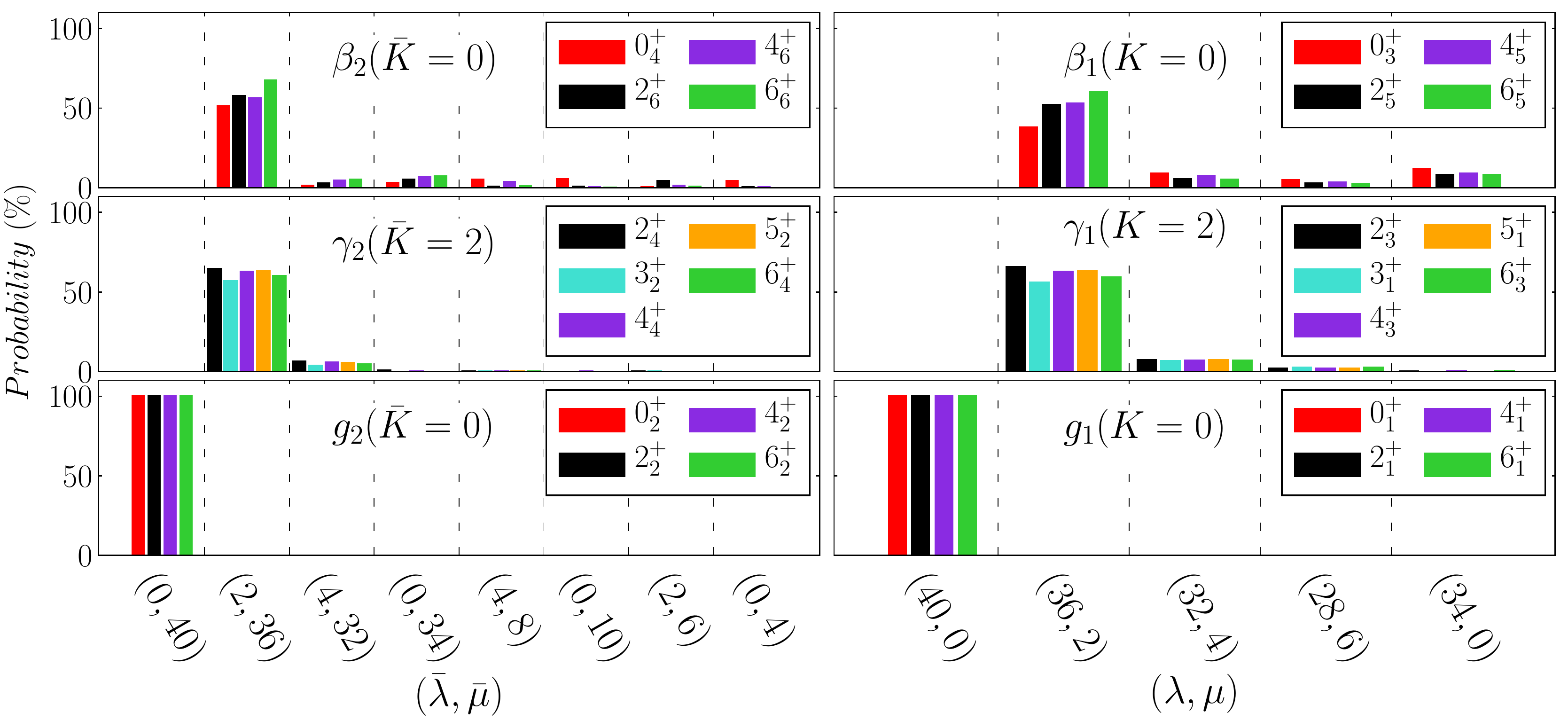}
\caption{
SU(3) $(\lambda,\mu)$- and $\bsu3$ $(\blam,\bmu)$-decompositions 
for members of the prolate ($g_1,\beta_1,\gamma_1$) 
and oblate ($g_2,\beta_2,\gamma_2$) bands, eigenstates of 
$\hat{H}'$ (\ref{HprimePO}) with parameters as in Fig.~4, 
resulting in prolate-oblate (P-O) shape coexistence. 
Shown are probabilities larger than 4\%.
States in the prolate ($g_1$) and oblate ($g_2$) ground bands 
are pure with respect to SU(3) and $\bsu3$, respectively.
In contrast, excited prolate and oblate bands are mixed, 
thus demonstrating the presence in the spectrum of 
SU(3)-PDS and $\bsu3$-PDS.
\label{fig5PO}}
\end{figure}

The members of the prolate and oblate ground-bands, 
Eq.~(\ref{HvanishPO}), 
are zero-energy eigenstates of $\hat{H}$ (\ref{HintPO}), 
with good SU(3) and $\bsu3$ symmetry, 
respectively. The Hamiltonian is invariant under a change of sign 
of the $s$-bosons, hence commutes with the ${\cal R}_{s}$ operator 
mentioned above. 
Consequently, all non-degenerate eigenstates of $\hat{H}$ 
have well-defined $s$-parity. 
This implies vanishing quadrupole moments for an $E2$ operator 
which is odd under such sign change.
To overcome this difficulty, we introduce a small $s$-parity 
breaking term ${\textstyle\alpha\hat{\theta}_2}$, Eq.~(\ref{theta2}), 
which contributes to $\tilde{E}(\beta,\gamma)$ a component 
${\textstyle\tilde{\alpha}(1+\beta^2)^{-2}[ 
(\beta^2\!-\!2)^2 \!+\! 2\beta^2(2 \!-\!2\sqrt{2}\beta\Gamma 
\!+\!\beta^2)]}$, 
with ${\textstyle\tilde{\alpha}=\alpha/(N-2)}$. 
The linear $\Gamma$-dependence distinguishes 
the two deformed minima and slightly lifts 
their degeneracy, as well as that of the normal modes~(\ref{d-modes}). 
Replacing ${\textstyle\hat{\theta}_2}$ 
by ${\textstyle\bar{\theta}_2 \!=\! 
-\hat{C}_{2}[\bsu3] + 2\hat{N}(2\hat{N}+3)}$, 
leads to similar effects but
interchanges the role of prolate and oblate bands. 
Identifying the collective part with $\hat{C}_2[{\rm SO(3)}]$, 
we arrive at the following complete Hamiltonian 
\ba
\hat{H}' &=& 
h_0\,P^{\dag}_0\hat{n}_sP_0 + h_2\,P^{\dag}_0\hat{n}_dP_0 
+\eta_3\,G^{\dag}_3\cdot\tilde{G}_3
+ \alpha\,\hat{\theta}_2 
+ \rho\,\hat{C}_2[\rm SO(3)] ~.
\label{HprimePO}
\ea 
The prolate $g_1$-band 
remains solvable with energy $E_{g1}(L)=\rho L(L+1)$. 
The oblate $g_2$-band experiences a slight shift of 
order ${\textstyle\tfrac{32}{9}\alpha N^2}$ and 
displays a rigid-rotor like spectrum. 
The SU(3) and $\bsu3$ decomposition in Fig.~\ref{fig5PO} demonstrates 
that these bands are pure DS basis states, with 
$(2N,0)$ and $(0,2N)$ character, respectively, 
while excited $\beta$ and $\gamma$ bands exhibit considerable mixing. 
The critical-point Hamiltonian thus has a subset of states with good SU(3) 
symmetry, a subset of states with good $\bsu3$ symmetry and all other states 
are mixed with respect to both SU(3) and $\bsu3$. These are precisely the 
defining ingredients of SU(3)-PDS coexisting with $\bsu3$-PDS.
\begin{figure}[t]
  \centering
\includegraphics[width=7cm]{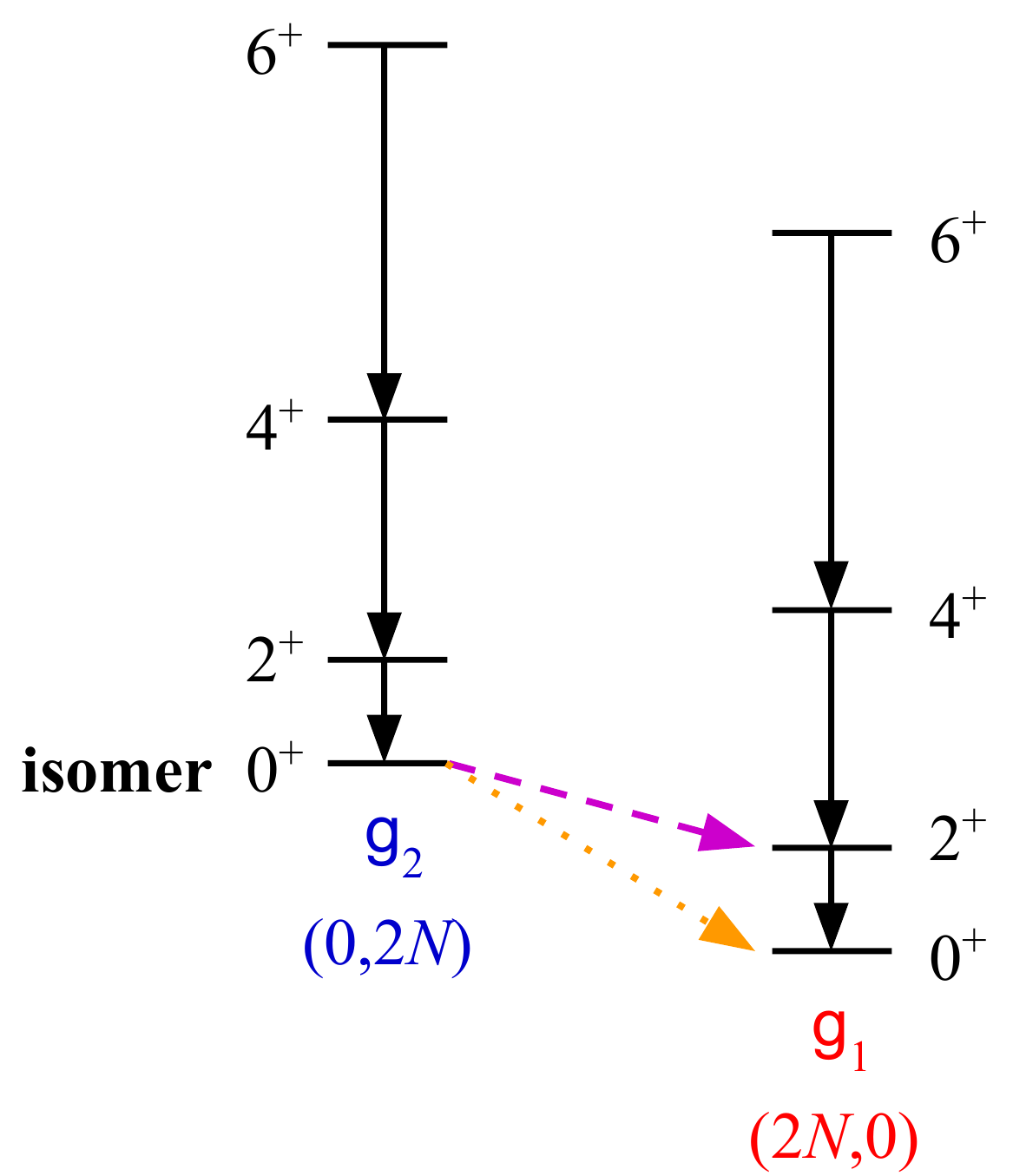}
\caption{\label{fig6PObe2}
Signatures of SU(3) and $\bsu3$ PDSs in P-O shape coexistence.
Strong intraband $E2$ transitions (solid lines) 
obey the analytic expression of Eq.~(\ref{be2}).
Retarded $E2$ (dashes lines) and $E0$ (dotted lines) decays identify 
isomeric states.}
\end{figure}

Since the wave functions for the members of the $g_1$ and $g_2$ bands 
are known, one can derive analytic expressions for their 
quadrupole moments and $E2$ transition rates. 
Considering the $E2$ operator 
$T(E2) = e_B\,\Pi^{(2)}$ with
\ba
\Pi^{(2)} = d^{\dag}s + s^{\dag}\tilde{d} ~,
\label{Pi2}
\ea
the quadrupole moments 
are found to have equal magnitudes and opposite signs, 
\ba
Q_L &=& 
{\textstyle
\mp e_B\sqrt{\frac{16\pi}{40}}\frac{L}{2L+3}
\frac{4(2N-L)(2N+L+1)}{3(2N-1)}} ~,
\label{quadmom}
\ea
where the minus (plus) sign corresponds to the prolate-$g_1$ (oblate-$g_2$) 
band. The $B(E2)$ values for intraband ($g_1\to g_1$, $g_2\to g_2$) 
transitions, 
\ba
&&B(E2; g_i,\, L+2\to g_i,\,L) = 
\nonumber\\
&&
\quad
\;\;
{\textstyle
e_{B}^2\,\frac{3(L+1)(L+2)}{2(2L+3)(2L+5)}
\frac{(4N-1)^2(2N-L)(2N+L+3)}{18(2N-1)^2}} ~,
\qquad\qquad
\label{be2}
\ea
are the same. These properties are ensured by the fact that 
${\cal R}_sT(E2){\cal R}_s^{-1} = -T(E2)$. Interband 
$(g_2\leftrightarrow g_1)$ 
$E2$ transitions, are extremely weak. This follows from the fact that 
the $L$-states of the $g_1$ and $g_2$ bands exhaust, respectively, 
the $(2N,0)$ and $(0,2N)$ irrep of SU(3) and $\bsu3$. 
$T(E2)$ contains a $(2,2)$ tensor under both algebras, 
hence can connect the $(2N,0)$ irrep of $g_1$ 
only with the $(2N-4,2)$ component in $g_2$ which, however, is 
vanishingly small. The selection rule $g_1\nleftrightarrow g_2$ 
is valid also for a more general $E2$ operator, 
obtained by including in it the operators 
$Q^{(2)}$~(\ref{Qsu3}) or $\bar{Q}^{(2)}$~(\ref{Qsu3bar}), 
since the latter, as generators, 
cannot mix different irreps of SU(3) or $\bsu3$. 
By similar arguments, $E0$ transitions in-between 
the $g_1$ and $g_2$ bands are extremely weak, 
since the relevant operator, 
$T(E0)\propto\hat{n}_d$, is a combination of $(0,0)$ and $(2,2)$ 
tensors under both algebras. Accordingly, 
the $L=0$ bandhead state of the higher ($g_2$) band, 
cannot decay by strong $E2$ or $E0$ transitions to the lower $g_1$ band, 
hence, as depicted schematically in Fig.~6, displays characteristic 
features of an isomeric state. In contrast to $g_1$ and $g_2$, excited 
$\beta$ and $\gamma$ bands are mixed, hence are connected by $E2$ transitions 
to these ground bands. 
Their quadrupole moments are found numerically to resemble, for large $N$, 
the collective model expression 
$Q(K,L) = \frac{3K^2-L(L+1)}{(L+1)2L+3)} q_{K}$,
with $q_{K}>0$ ($q_{K}<0$) for prolate (oblate) bands.

\section{Triple spherical-prolate-oblate coexistence: 
U(5)-SU(3)-$\mathbf{\overline{SU(3)}}$ PDS}
\label{SPOshapes}

Nuclei can accommodate more than two shapes simultaneously, 
as encountered in the triple coexistence of spherical, prolate and 
oblate shapes. The relevant DS limits for the latter configurations 
are~\cite{ibm},
\bsub
\ba
&&{\rm U(6)\supset U(5)\supset SO(5)\supset SO(3)} \;\;\quad
\ket{N,\, n_d,\,\tau,\,n_{\Delta},\,L} ~.\quad 
\label{U5tri}
\\
&&{\rm U(6)\supset SU(3)\supset SO(3)} \;\,\,\qquad\quad\quad
\ket{N,\, (\lambda,\mu),\,K,\, L} ~,\quad 
\label{SU3tri}
\\
&&{\rm U(6)\supset \bsu3\supset SO(3)} \;\,\,\qquad\quad\quad
\ket{N,\, (\blam,\bmu),\,\bar{K},\, L} ~.\quad
\label{SU3bartri}
\ea
\label{u5su3su3bchains}
\esub
Properties of the above U(5), SU(3) and $\bsu3$ DS chains 
were discussed in Sections 4-5.
The intrinsic part of the critical-point Hamiltonian is now 
required to satisfy three conditions
\bsub
\ba
\hat{H}\ket{N,\, n_d=0,\,\tau=0,\,L=0} &=& 0 ~,
\label{nd0tri}
\\
\hat{H}\ket{N,\, (\lambda,\mu)=(2N,0),\,K=0,\, L} &=& 0 ~,
\label{2N0tri}
\\
\hat{H}\ket{N,\, (\blam,\bmu)=(0,2N),\,\bar{K}=0,\, L} &=& 0 ~.
\label{02Ntri}
\ea
\label{vanishtri}
\esub
Equivalently, $\hat{H}$ annihilates the 
spherical intrinsic state of 
Eq.~(\ref{int-state}) with $\beta=0$, 
which is the single basis state in the U(5) irrep $n_d=0$, and 
the deformed intrinsic states with $(\beta\!=\!\sqrt{2},\gamma\!=\!0)$ 
and $(\beta\!=\!-\sqrt{2},\gamma\!=\!0)$, which are the lowest and 
highest-weight vectors in the irreps $(2N,0)$ and $(0,2N)$ 
of SU(3) and $\bsu3$, respectively. 
The resulting Hamiltonian is found to be that of 
Eq.~(\ref{HintPO}) with $h_0=0$~\cite{LevDek16}, 
\ba
\hat{H} = 
h_2\,P^{\dag}_0\hat{n}_dP_0 +\eta_3\,G^{\dag}_3\cdot\tilde{G}_3 ~.
\label{HintSPO}
\ea
\begin{figure}[t]
  \centering
\includegraphics[width=16cm]{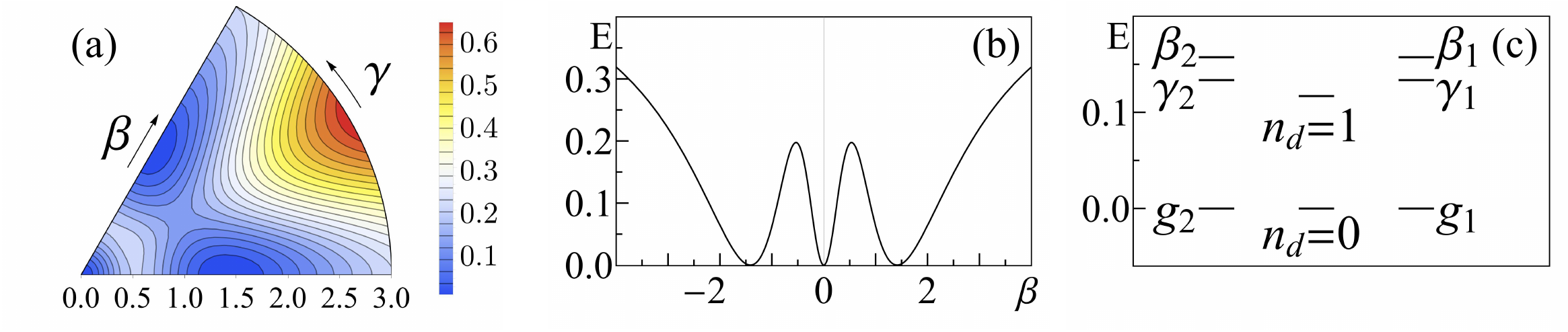}
  \caption{
Spherical-prolate-oblate (S-P-O) shape coexistence.
(a)~Contour plots of the energy surface~(\ref{surfacetri}),   
(b)~$\gamma\!=\!0$ sections, and
(c)~bandhead spectrum, for the Hamiltonian $\hat{H}'$~(\ref{HprimeSPO})
with parameters $h_2\!=\!0.5,\,\eta_3\!=\!0.567,\,\alpha\!=\!0.018,
\,\rho=0$ and $N\!=\!20$.
\label{fig7SPO}}
\end{figure}
The corresponding energy surface, 
$E_{N}(\beta,\gamma) = N(N-1)(N-2)\tilde{E}(\beta,\gamma)$, 
is given by
\ba
\tilde{E}(\beta,\gamma) = 
\beta^2 [\,h_2(\beta^2-2)^2
+\eta_3 \beta^4\sin^2(3\gamma)\,](1+\beta^2)^{-3} ~.
\label{surfacetri}
\ea
For $h_2,\eta_3\geq 0$, 
$\hat{H}$ is positive definite and 
$\tilde{E}(\beta,\gamma)$ has three degenerate global minima: 
$\beta=0$, $(\beta=\sqrt{2},\gamma=0)$ and $(\beta=\sqrt{2},\gamma=\pi/3)$ 
[or equivalently $(\beta=-\sqrt{2},\gamma=0)$], at $\tilde{E}=0$. 
Additional extremal points include 
(i)~a saddle point: $[\bs^2 = \frac{2}{7}, \gamma=0,\pi/3]$, 
at $\tilde{E}=\frac{32}{81}h_2$.
(ii)~A~local maximum and a saddle point: $[\bss^2,\gamma=\pi/6]$, 
at $\tilde{E}= \frac{4}{3}h_2(1+\bss^2)^{-2}\bss^2(2-\bss^2)$, 
where $\bss^2$ is a solution of 
${\textstyle(7h_2+3\eta_3)\bss^4 -16h_2\bss^2 + 4h_2 = 0}$. 
The saddle points, when they exist, support 
a barrier separating the various minima, as seen in Fig.~7. 
The normal modes involve quadrupole vibrations about the spherical minimum 
with frequency $\epsilon$ alongside $\beta$ and $\gamma$ vibrations 
about the deformed prolate and oblate minima 
with frequencies $\epsilon_{\beta i}$ and $\epsilon_{\gamma i}$,
\bsub
\ba
&&\epsilon = 4h_2N^2 ~,
\\
&&\epsilon_{\beta 1}=\epsilon_{\beta 2} 
= \frac{16}{3}h_2N^2 ~,
\\
&&\epsilon_{\gamma 1}=\epsilon_{\gamma 2} = 4\eta_3N^2 ~.
\ea
\label{sd-modes}
\esub
The spherical modes are seen to have a lower frequency than the 
$\beta$ modes, $\epsilon=\tfrac{3}{4}\epsilon_{\beta i}$. 
Figs.~(7a), (7b) and (7c) show
$\tilde{E}(\beta,\gamma)$, $\tilde{E}(\beta,\gamma\!=\!0)$ 
and the normal-mode spectrum 
with parameters ensuring spherical-prolate-oblate (S-P-O) minima. 
\begin{figure}[t]
  \centering
\includegraphics[width=16cm]{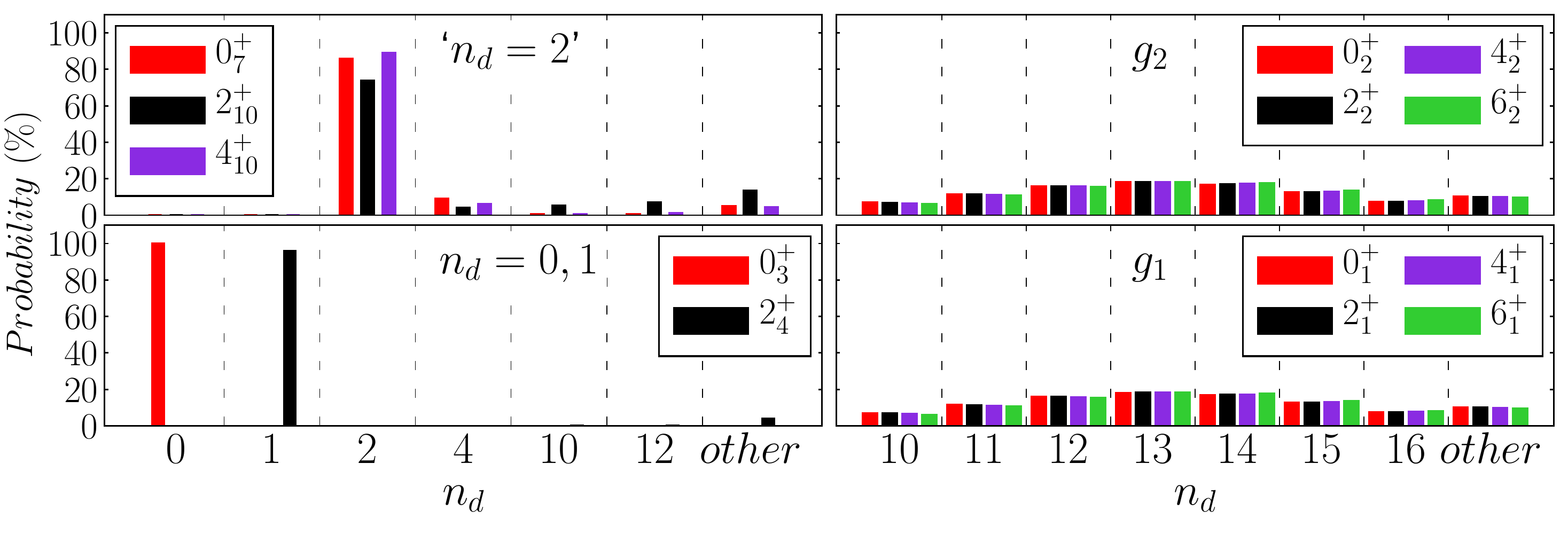}
\caption{
U(5) $n_d$-decomposition for spherical states (left panels)
and for members of the deformed prolate ($g_1$) and oblate ($g_2$) 
ground bands (right panels), 
eigenstates of 
$\hat{H}'$ (\ref{HprimeSPO}) with parameters as in Fig.~\ref{fig7SPO}, 
resulting in spherical-prolate-oblate (S-P-O) shape coexistence.
The column `other' depicts a sum of probabilities, each less than 5\%. 
The spherical states are dominated by a single $n_d$ component, in marked 
contrast to the deformed states, thus signaling 
the presence in the spectrum of U(5)-PDS.
\label{fig8SPO}}
\end{figure}

For the same arguments as in the analysis of prolate-oblate 
shape coexistence in Section~5, the complete Hamiltonian is taken to be
\ba
\hat{H}' &=& 
h_2\,P^{\dag}_0\hat{n}_dP_0 +\eta_3\,G^{\dag}_3\cdot\tilde{G}_3 
+ \alpha\,\hat{\theta}_2 
+ \rho\,\hat{C}_2[\rm SO(3)] ~.
\label{HprimeSPO}
\ea 
The deformed bands show similar rigid-rotor structure 
as in the prolate-oblate case. 
In particular, the prolate $g_1$-band and oblate $g_2$-band have 
good SU(3) and $\bsu3$ symmetry, respectively, 
while excited $\beta$ and $\gamma$ bands 
exhibit considerable mixing, with similar decompositions as in Fig.~5.
A new aspect in the present S-P-O case, 
is the simultaneous occurrence in the spectrum [see Fig.~7(c)] 
of spherical type of states, whose wave functions are dominated by 
a single $n_d$ component. 
As shown in the left panels of Fig.~8, 
the lowest spherical $L=0^{+}_3$ state is a pure $n_d=0$ state,
which is the solvable U(5) basis state of Eq.~(\ref{nd0tri}). 
The $L=2^{+}_4$ state is almost pure, with a probability of 
96.1\% for the $n_d=1$ component. The origin of its high degree of purity 
can be traced to the relation
\ba
&&
\hat{H}\ket{N,n_d\!=\!\tau\!=\!1,L\!=\!2} = \qquad
\nonumber\\ 
&&\qquad
{\textstyle
h_2\,4(N-1)(N-2)\left [\ket{N,n_d\!=\!\tau\!=\!1,L\!=\!2} 
- \sqrt{\frac{7}{2(N-1)(N-2)}}\,
\ket{N,n_d\!=\!\tau\!=\!L\!=\!2}\right ]} ~,\qquad
\ea
which shows that the U(5) basis state $\ket{N,n_d\!=\!\tau\!=\!1,L\!=\!2}$ 
approaches the status of an eigenstate 
for large $N$, with corrections of order $1/N$.
Higher spherical type of states $(L=0^{+}_7,\,2^{+}_{10},\,4^{+}_{10})$ 
have a pronounced ($\sim$~80\%) $n_d\!=\!2$ component. This structure 
should be contrasted with the U(5) decomposition of deformed states 
({\it e.g.}, those belonging to the $g_1$ and $g_2$ bands) 
which, as shown in the right panels of Fig.~8, 
have a broad $n_d$-distribution. 
The purity of selected sets of states 
with respect to SU(3), $\bsu3$ and U(5) as demonstrated in Figs.~5 and~8, 
in the presence of other mixed states, 
are the hallmarks of coexisting SU(3)-PDS, $\bsu3$-PDS and U(5)-PDS. 
It is remarkable that a simple Hamiltonian, as in Eq.~(\ref{HprimeSPO}), 
can accommodate simultaneously several incompatible symmetries in 
a segment of its spectrum.
\begin{figure}[t]
  \centering
\includegraphics[width=7cm]{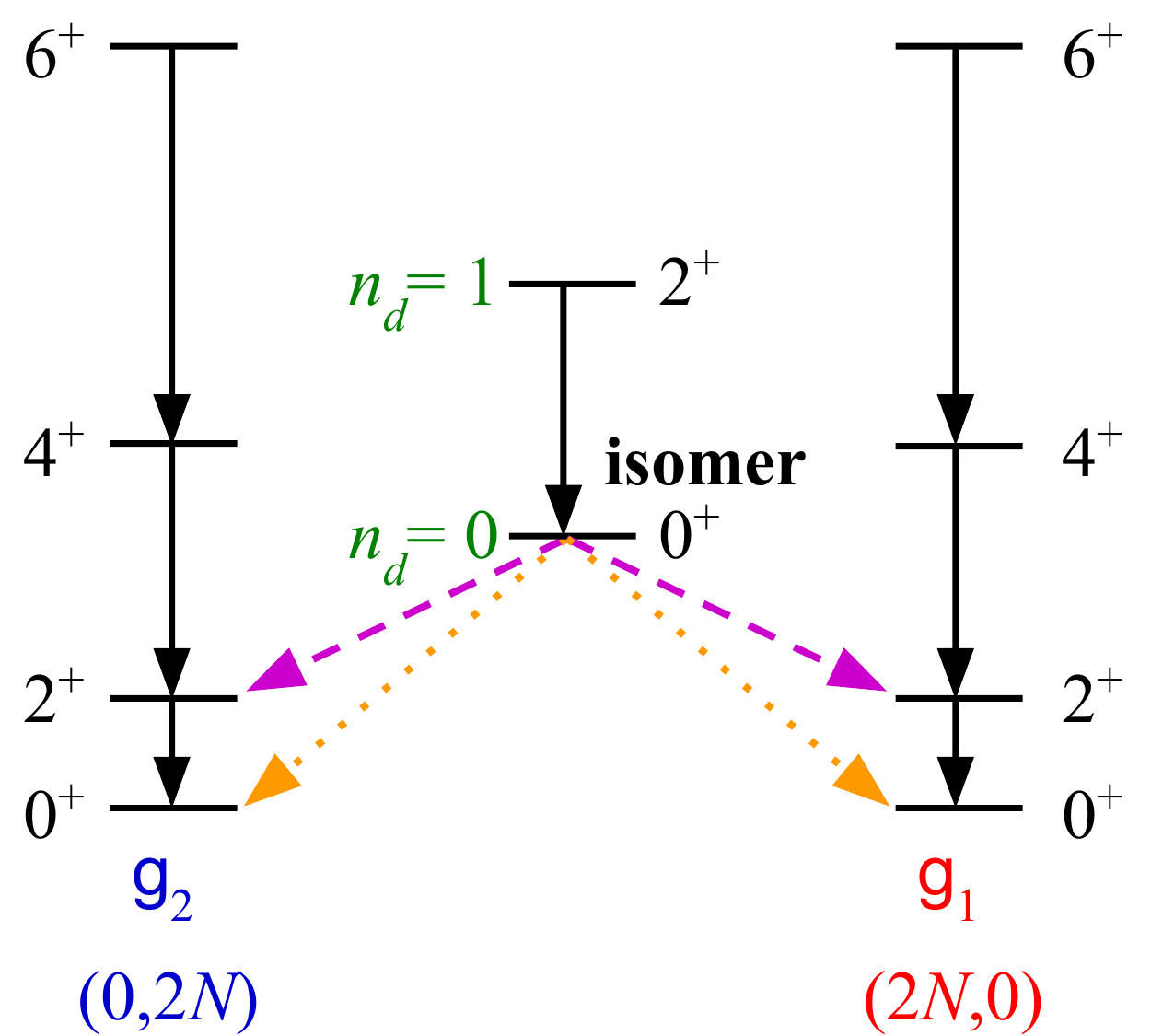}
\caption{\label{fig9SPObe2}
Signatures of U(5), SU(3) and $\bsu3$ PDSs in S-P-O shape coexistence.
Strong intraband $E2$ transitions (solid lines) 
obey the analytic expressions of Eqs.~(\ref{be2}) and~(\ref{be2nd}). 
Retarded $E2$ (dashes lines) and $E0$ (dotted lines) decays 
identify isomeric states.}
\end{figure}

Considering the same $E2$ operator $T(E2)=e_B\Pi^{(2)}$, Eq.~(\ref{Pi2}), 
as in the prolate-oblate case of Section~5, 
the quadrupole moments of 
states in the solvable $g_1$ and $g_2$ bands and 
intraband ($g_1\to g_1$, $g_2\to g_2$) $E2$ rates, 
obey the analytic expressions of 
Eqs.~(\ref{quadmom}) and (\ref{be2}), respectively. 
The same selection rules depicted in Fig.~6, resulting in retarded 
$E2$ and $E0$  
interband  ($g_2\to g_1$) decays, are still valid. 
Furthermore, in the current S-P-O case, 
since $T(E2)$ obeys the selection rule $\Delta n_d\!=\!\pm 1$, the 
spherical states, $(n_d\!=\!L\!=\!0)$ and $(n_d\!=\!1,L\!=\!2)$, 
have no quadrupole moment and the $B(E2)$ value for their 
connecting transition, obeys the U(5)-DS expression~\cite{ibm}
\ba
B(E2; n_d=1,L=2\to n_d=0,L=0) = e_{B}^2N ~.\quad
\label{be2nd}
\ea
These spherical states have very weak $E2$ transitions to the 
deformed ground bands, because they exhaust the $(n_d\!=\!0,1)$ irreps 
of U(5), and the $n_d\!=\!2$ component in the ($L\!=\!0,2,4$) states 
of the $g_1$ and $g_2$ bands is 
extremely small, of order $N^33^{-N}$, as can be inferred from 
Eq.~(\ref{g1u5expan}). 
There are also no $E0$ transitions involving these spherical states, 
since $T(E0)$ is diagonal in $n_d$.

In the above analysis the spherical and deformed minima were assumed 
to be near degenerate. If the spherical minimum is only 
local, one can use the Hamiltonian of Eq.~(\ref{HprimePO}) 
with the condition $h_2>4h_0$, 
for which the spherical ground state $(n_d=L=0)$ 
experiences a shift of order $4h_0N^3$.
Similarly, if the deformed minima are only local, 
adding an $\epsilon\hat{n}_d$ term to $\hat{H}'$ (\ref{HprimeSPO}), 
will leave the $n_d=0$ spherical ground state unchanged, but will shift 
the prolate and oblate bands to higher energy of order $2\epsilon N/3$. 
In both scenarios, the lowest $L=0$ state of the non-yrast configuration 
will exhibit retarded $E2$ and $E0$ decays, hence will have the attributes 
of an isomer state, as depicted schematically in Fig.~9

\section{Spherical and $\gamma$-unstable deformed shape coexistence: 
U(5)-SO(6) PDS}
\label{SGshapes}

The $\gamma$ degree of freedom and triaxiality can play an 
important role in the occurrence of multiple shapes in nuclei~\cite{Ayan16}. 
In the present section, we examine the coexistence of a spherical shape 
and a particular type of non-axial deformed shape, which is $\gamma$-soft. 
The relevant DS chains for such configurations are~\cite{ibm},
\bsub
\ba
&&
{\rm U(6)\supset U(5)\supset SO(5)\supset SO(3)} \;\;\;\;\quad\quad
\ket{N,\, n_d,\,\tau,\,n_{\Delta},\, L} ~,\quad 
\label{u5SG}
\\
&&
{\rm U(6)\supset SO(6)\supset SO(5)\supset SO(3)} \;\;\quad\quad
\ket{N,\, \sigma,\,\tau,\,n_{\Delta},\, L} ~.\quad 
\label{o6SG}
\ea
\label{u5o6}
\esub
The U(5)-DS limit~(\ref{u5SG}), appropriate to a spherical shape, 
was discussed in Section~\ref{SPshapes}. 
The SO(6)-DS limit~(\ref{o6SG}) is appropriate to the dynamics of 
a $\gamma$-unstable deformed shape. 
For a given U(6) irrep $N$, the allowed SO(6) and SO(5) irreps are 
$\sigma\!=\!N,\,N-2,\dots 0$ or $1$, and  
$\tau=0,\,1,\,\ldots \sigma$, respectively. 
The ${\rm SO(5)}\supset {\rm SO(3)}$ reduction is the same as in the 
U(5)-DS chain. 
The basis states are eigenstates of the Casimir operator 
$\hat{C}_{2}[{\rm SO(6)}] \!=\! 2\Pi^{(2)}\cdot \Pi^{(2)} \!+\! 
\hat{C}_{2}[{\rm SO(5)}]$ 
with eigenvalues $\sigma(\sigma+4)$.  
The generators of SO(6) are the angular momentum, octupole and 
quadrupole operators, 
$L^{(1)},\,(d^{\dag}\tilde{d})^{(3)}$ and $\Pi^{(2)}$, Eq.~(\ref{Pi2}). 
The SO(6)-DS spectrum resembles that of a $\gamma$-unstable deformed 
rotovibrator, composed of SO(6) $\sigma$-multiplets forming rotational 
bands, with $\tau(\tau+3)$ and $L(L+1)$ splitting generated 
by $\hat{C}_{2}[{\rm SO(5)}]$ and 
$\hat{C}_{2}[{\rm SO(3)}]$, respectively. 
The lowest irrep $\sigma\!=\!N$ contains the ground ($g$) band  
of a $\gamma$-unstable deformed nucleus. 
The first excited irrep $\sigma\!=\!N-2$ contains the $\beta$-band. 
The lowest members in each band have quantum numbers 
$(\tau=0,\, L=0)$, $(\tau=1,\, L=2)$, 
$(\tau=2,\, L=2,4)$ and $(\tau=3,\, L=0,3,4,6)$. 

In discussing properties of the SO(6)-DS spectrum, it is convenient 
to subtract from $\hat{C}_{2}[{\rm O(6))}]$ the ground-state energy, and 
consider the following positive-definite term
\ba
R^{\dag}_0R_0 =
-\hat C_{2}[{\rm SO(6)}] + \hat{N} (\hat{N}+4) ~.
\label{R0R0}
\ea 
The SO(6) basis states 
$\ket{N,\, \sigma,\, \tau,\, n_{\Delta},\, L}$, Eq.~(\ref{o6SG}),
are then eigenstates of $R^{\dag}_0R_0$ with eigenvalues 
$(N-\sigma)(N+\sigma+4)$, and the ground 
band with $\sigma=N$ occurs at zero energy. 
The two-boson pair operator 
\ba
R^{\dagger}_{0} &=& d^{\dagger}\cdot d^{\dagger} - (s^{\dagger})^2 ~,
\label{R0}
\ea
is a scalar with respect to SO(6) and satisfies 
\ba
R_{0}\,\vert N,\sigma=N,\tau, n_{\Delta}, L\rangle &=& 0 ~.
\label{R0vanish}
\ea
This operator corresponds to 
$\hat{T}_{\alpha}$ of Eq.~(\ref{Talpha}) and, as shown below, it plays 
a central role in the construction of Hamiltonians with SO(6)-PDS.
\begin{figure}[t]
  \centering
\includegraphics[width=16cm]{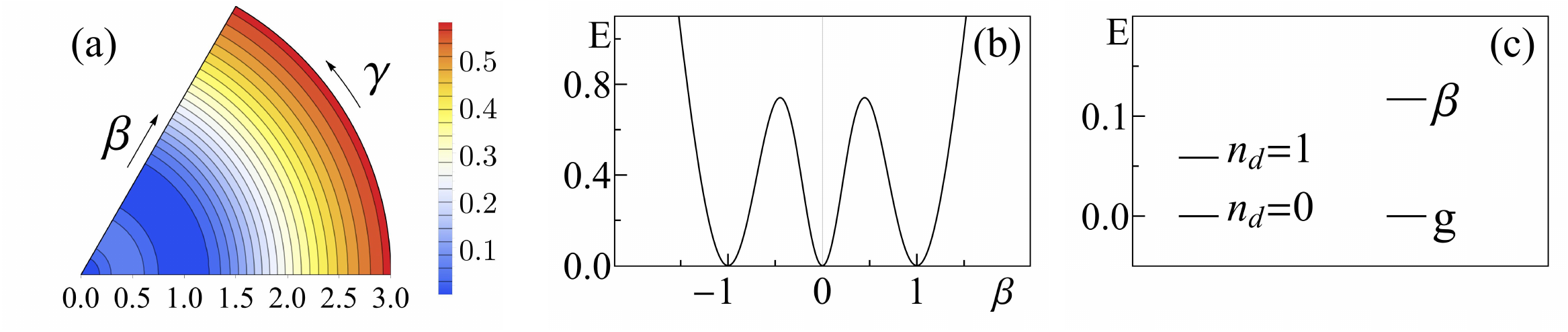}
  \caption{
Spherical and $\gamma$-unstable deformed (S-G) shape coexistence.
(a)~Contour plots of the $\gamma$-independent 
energy surface~(\ref{surfaceSG}),   
(b)~$\gamma\!=\!0$ sections, and
(c)~bandhead spectrum, for the Hamiltonian $\hat{H}'$~(\ref{HprimeSG})
with parameters $r_2\!=\!1,\,\rho_5\!=\!\rho_3=0$, and $N\!=\!20$.
\label{fig10SG}}
\end{figure}

The phase transition between spherical and $\gamma$-unstable deformed 
shapes, has been previously studied by varying a control parameter 
in an IBM Hamiltonian mixing the U(5) and SO(6) Casimir 
operators~\cite{diep80,CejJolCas10,Iac11,LevGin03}. 
However, the latter employed one- and two-body terms, hence 
the resulting quantum phase transition is of second order, 
where one minimum evolves continuously to the second minimum. 
To allow for a first-order quantum phase transition, involving 
coexisting shapes, 
we consider an Hamiltonian with cubic terms which retains the virtues 
of U(5) and SO(6) DSs for the spherical ground state and the 
$\gamma$-unstable deformed ground band, respectively. 
Following the procedure outlined in Eq.~(\ref{bases12}),
the intrinsic part of the 
critical-point Hamiltonian is required to satisfy
\bsub
\ba
\hat{H}\ket{N,\, \sigma=N,\,\tau,\, L} &=& 0 
\qquad\qquad \tau=0,1,2,\ldots,N
\label{sigmaN}
\\
\hat{H}\ket{N,\, n_d=0,\,\tau=0,\,L=0} &=& 0 ~.
\label{nd0SG}
\ea
\label{o6u5solv}
\esub 
Equivalently, $\hat{H}$ annihilates both the deformed intrinsic state 
of Eq.~(\ref{int-state}) with $(\beta=1,\gamma\,{\rm arbitrary})$,
which is the lowest weight vector in the SO(6) irrep
$\sigma=N$, and the spherical intrinsic state with $\beta=0$, 
which is the single basis state in the U(5) irrep $n_d=0$. 
The resulting Hamiltonian is found to be 
\ba
\hat{H} = 
r_2\,R^{\dag}_0\hat{n}_dR_0 ~,
\label{HintSG}
\ea
where $R^{\dag}_0$ is given in Eq.~(\ref{R0}). 
The energy surface, 
$E_{N}(\beta,\gamma) = N(N-1)(N-2)\tilde{E}(\beta,\gamma)$, 
is given by
\ba
\tilde{E}(\beta) = 
r_2\beta^2(\beta^2-1)^2 (1+\beta^2)^{-3} ~.
\label{surfaceSG}
\ea
The surface is an even sextic function of $\beta$ and is independent 
of $\gamma$, in accord with the SO(5) symmetry of the Hamiltonian.
It has the form
$\tilde{E}(\beta) = 
(1+\beta^2)^{-3}[A\beta^6 + D\beta^4+ F\beta^2]$, 
with coefficients $A \!=\! F \!=\! r_2,\, D \!=\! -2r_2$, 
satisfying $D^2=4AF$. Such a topology 
necessitates the presence of 
cubic terms in the Hamiltonian, 
and the latter condition ensures that the surface supports 
two degenerate extrema, spherical and deformed.
For $r_2 > 0$, 
$\hat{H}$ is positive definite and 
$\tilde{E}(\beta)$ has two degenerate global minima, 
$\beta=0$ and $\beta^2=1$, at $\tilde{E}=0$.
A local maximum at $\bs^2=\frac{1}{5}$ creates a barrier 
of height $\tilde{E} =\frac{2}{27}r_2$, separating the two minima, 
as seen in Fig.~10. 
For large $N$, the normal modes shown schematically in Fig.~(10c), 
involve $\beta$ vibrations about the deformed minima, with frequency 
$\epsilon_{\beta}$, and quadrupole vibrations about the spherical minimum, 
with frequency $\epsilon$, respectively, 
\bsub
\ba
\epsilon_{\beta} &=& 2r_2\,N^2 ~,
\\
\epsilon &=& r_2\,N^2 ~.
\label{sg-modes}
\ea
\esub
Interestingly, the $\beta$ mode has twice the energy of the spherical 
modes, $\epsilon_{\beta} = 2\epsilon$, compared to equal energies encountered
in the case of spherical-prolate coexistence [see 
Eqs.~(\ref{betmodes}) and (\ref{sphermodes})].
\begin{figure}[t]
  \centering
\includegraphics[width=15cm]{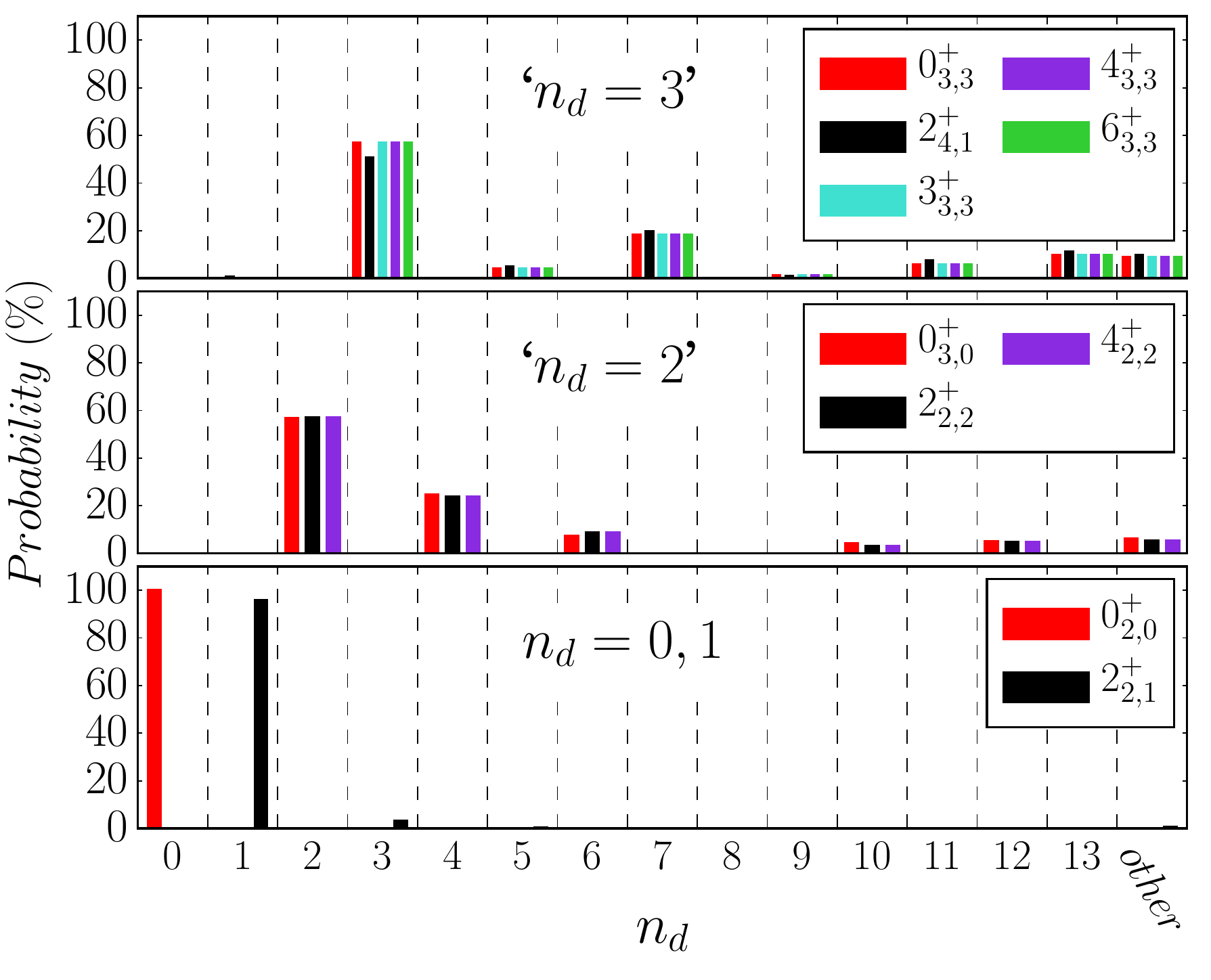}
\caption{
\label{fig11SG}
U(5) $n_d$-decomposition for spherical states, 
eigenstates of 
the Hamiltonian $\hat{H}'$~(\ref{HprimeSG}) 
with parameters as in Fig.~10. 
The column `other' depicts a sum of probabilities, each less than~5\%.} 
\end{figure}
\begin{figure}[t]
  \centering
\includegraphics[width=15cm]{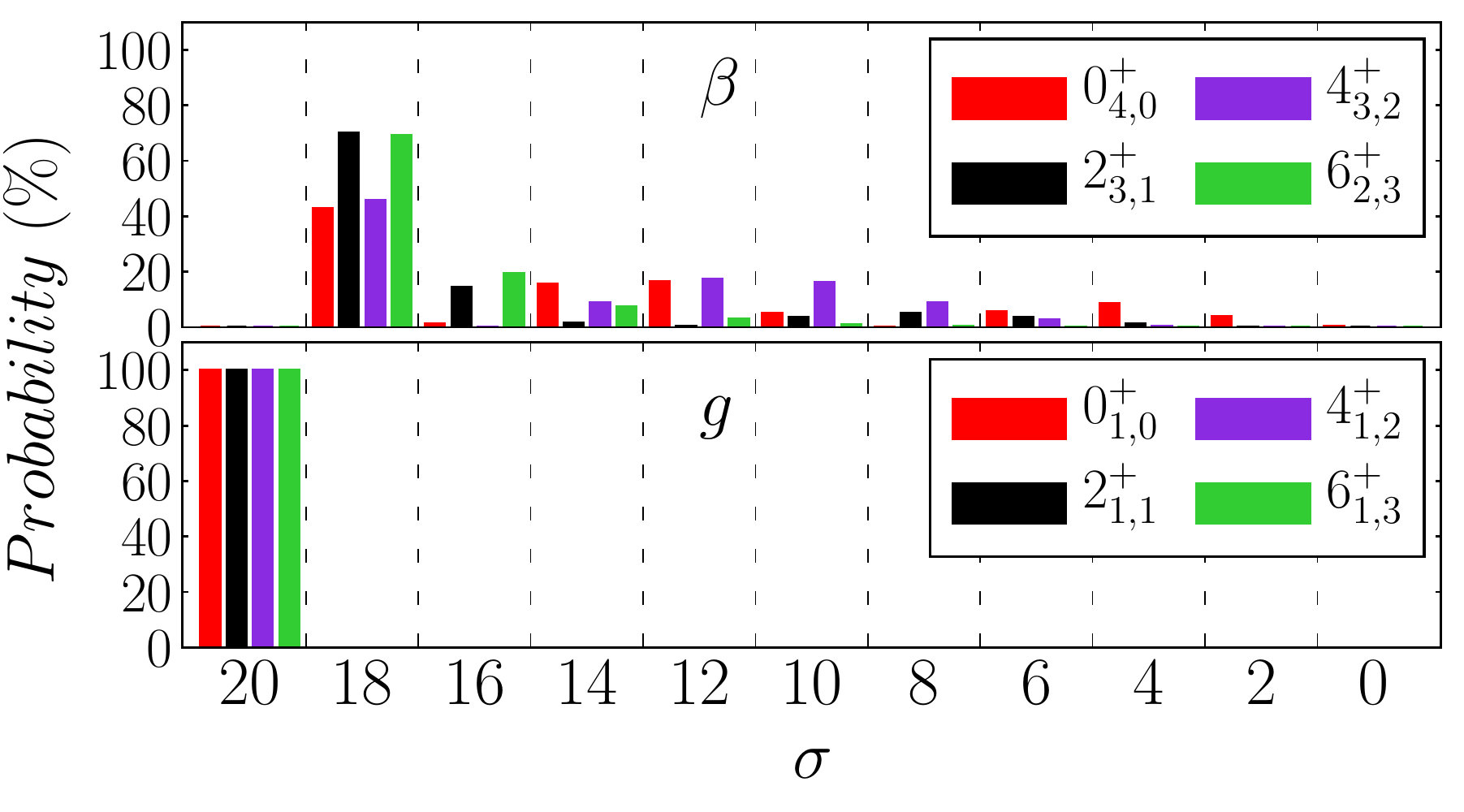}
\caption{
\label{fig12SG}
SO(6) $\sigma$-decomposition
for members of the deformed ground ($g$) 
and $\beta$ bands, 
eigenstates of 
the Hamiltonian $\hat{H}'$~(\ref{HprimeSG}) 
with parameters as in Fig.~10, 
resulting in S-G shape coexistence.} 
\end{figure}

Identifying the collective part of the Hamiltonian with the 
Casimir operators of the common ${\rm SO(5)\supset SO(3)}$ segment 
of the chains~(\ref{u5o6}), 
we arrive at the following complete Hamiltonian 
\ba
\hat{H}' &=& r_2\,R^{\dag}_0\hat{n}_dR_0
+ \rho_5\,\hat{C}_2[\rm SO(5)] + \rho_3\,\hat{C}_2[\rm SO(3)] ~.
\label{HprimeSG}
\ea 
The added rotational terms generate an exact 
$\rho_5\tau(\tau+3)+\rho_3L(L+1)$ splitting without 
affecting the wave functions.
In particular, the solvable subset of 
eigenstates, Eq.~(\ref{o6u5solv}), remain intact.
Since both SO(5) and SO(3) are preserved by the Hamiltonian, 
its eigenstates have good $(\tau,L)$ quantum numbers and can be labeled 
as $L^{+}_{i,\tau}$, where the ordinal number $i$ enumerates the occurrences 
of states with the same $(\tau,L)$ with increasing energy.
The nature of the Hamiltonian eigenstates can be inferred from the 
probability distributions, ${\textstyle P^{(N,\tau,L)}_{n_d} = 
\vert C_{n_d}^{(N,\tau,L)}\vert^2}$ and 
${\textstyle P^{(N,\tau,L)}_{\sigma} = 
\vert C_{\sigma}^{(N,\tau,L)}\vert^2}$, obtained from their expansion 
coefficients in the U(5) and SO(6) bases~(\ref{u5o6}). 
In general, the low lying spectrum of $\hat{H}'$~(\ref{HprimeSG}) 
exhibits two distinct classes of states. The first class consists 
of ($\tau,L$) states arranged in $n_d$-multiplets of a spherical vibrator.
Fig.~11 shows the U(5) $n_d$-decomposition 
of such spherical states, characterized by a narrow $n_d$-distribution. 
The lowest spherical state, $L=0^{+}_{2,0}$, is the solvable 
U(5) state of Eq.~(\ref{nd0SG}) with U(5) quantum number $n_d=0$. 
The $L=2^{+}_{2,1}$ state has $n_d=1$ to a good approximation.
Its high purity can be traced to the relation
\ba
&&\hat{H}\ket{N,n_d\!=\!\tau\!=\!1,L\!=\!2} =
\nonumber\\ 
&&\qquad\quad
{\textstyle
r_2(N\!-\!1)(N-2)\left [\ket{N,n_d\!=\!\tau\!=\!1,L\!=\!2} 
-\sqrt{\frac{14}{(N\!-\!1)(N\!-\!2)}}\,
\ket{N,n_d\!=\!\tau\!=\!L=\!2}\right ]} ~,\;\qquad
\ea
which shows 
that the U(5)-basis state $\ket{N,n_d\!=\!\tau\!=\!1,L\!=\!2}$ 
is almost an eigenstate for large $N$, with corrections of order $1/N$. 
The upper panels of Fig.~11 display the next spherical-type of 
multiplets ($L=0^{+}_{3,0},\,2^{+}_{2,2},\,4^{+}_{2,2}$)
and ($L=6^{+}_{3,3},\,4^{+}_{3,3},\,3^{+}_{3,3},\,0^{+}_{3,3},\,2^{+}_{4,1}$), 
which have a somewhat less pronounced (60\%)
single $n_d$-component, with $n_d=2$ and $n_d=3$, respectively.
\begin{figure}[t]
  \centering
\includegraphics[width=15cm]{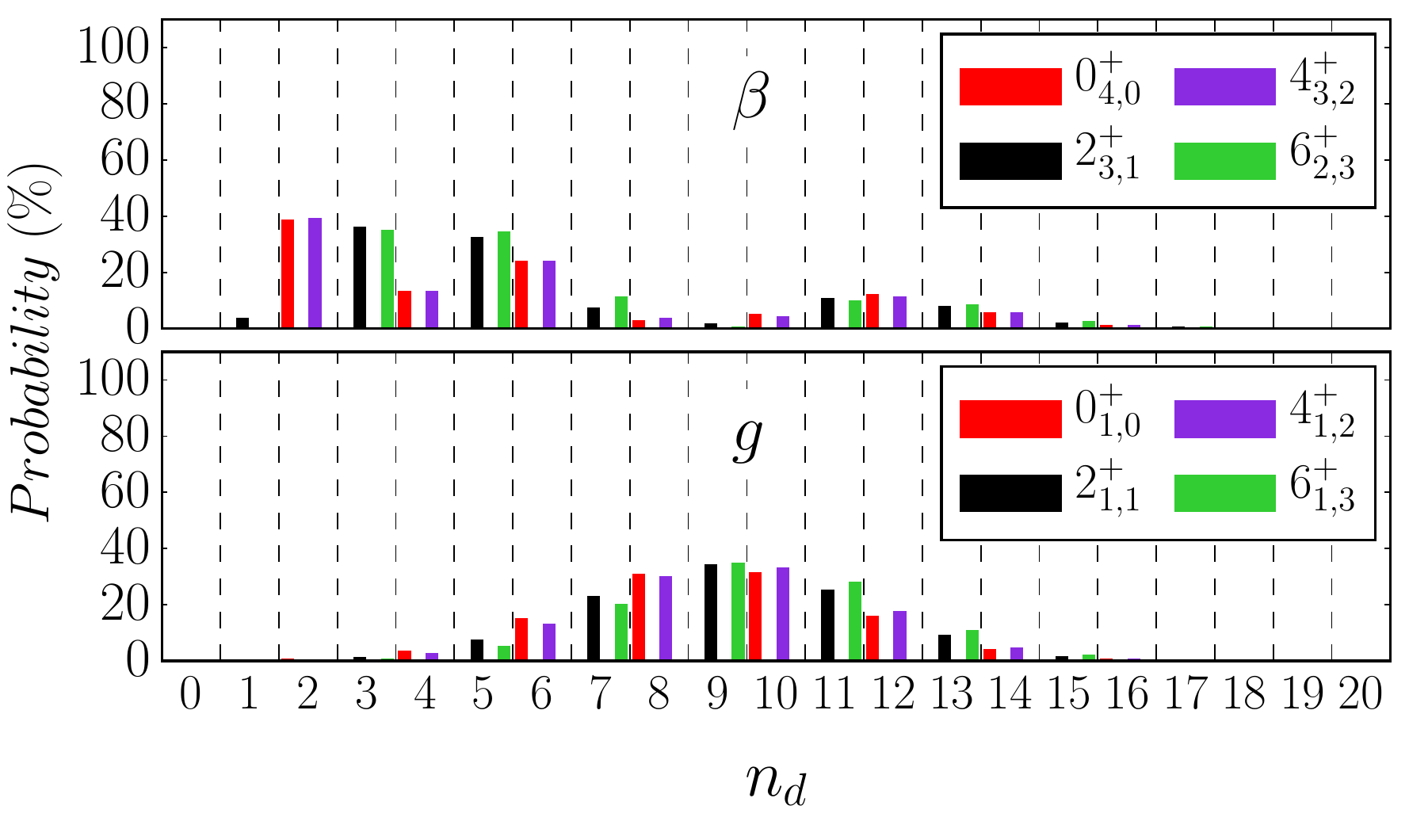}
\caption{
U(5) $n_d$-decomposition
for members of the deformed ground ($g$) 
and $\beta$ bands, eigenstates of 
the Hamiltonian $\hat{H}'$~(\ref{HprimeSG}) 
with parameters as in Fig.~10, 
resulting in spherical and $\gamma$-unstable deformed (S-G) 
shape coexistence. The results of Figs.~11-13, 
demonstrate the presence in the spectrum of U(5)-PDS and SO(6)-PDS.
\label{fig13SG}} 
\end{figure}

A second class consists of ($\tau,L$) states arranged in 
bands of a $\gamma$-unstable deformed rotor. 
The SO(6) $\sigma$-decomposition of such states, in selected bands, 
are shown in Fig.~12.
The ground band is seen to be pure with $\sigma=N$ SO(6) character, 
and coincides with the solvable band of Eq.~(\ref{sigmaN}).
In contrast, the non-solvable $\beta$-band (and higher $\beta^n$-bands) 
show considerable SO(6)-mixing. The deformed nature of these SO(5)-rotational 
states is manifested in their broad $n_d$-distribution, shown in Fig.~13.
This is explicitly evident in the following expansion 
of the SO(6) ground band wave functions in the U(5) basis,
\bsub
\ba
&& \ket{N,\sigma=N,\tau, n_{\Delta}, L} =
\sum_{n_d}\frac{1}{2}
\left [1 + (-1)^{n_d-\tau}\right ]\, \theta_{n_d}^{(N,\tau)}
\ket{N,n_d,\tau,n_{\Delta}, L} ~,\qquad\;\;\\
&&\theta_{n_d}^{(N,\tau)} =
\left[\frac{(N-\tau)!(N+\tau+3)!}
{2^{N+1}(N+1)!(N-n_d)!(n_d -\tau)!!(n_d+\tau+3)!!}\right]^{1/2} ~,
\qquad
\ea
\label{projnd}
\esub
which shows that for large $N$, the probability of each 
individual $n_d$ component is exponentially small. 
The above analysis demonstrates that although 
the critical-point Hamiltonian~(\ref{HprimeSG}) 
is not invariant under U(5) nor SO(6), 
some of its eigenstates have good U(5) symmetry, 
some have good SO(6) symmetry 
and all other states are mixed with respect to both U(5) and SO(6). 
These are precisely the defining attributes of U(5)-PDS coexisting 
with SO(6)-PDS.

Since the wave functions for the solvable states, 
Eqs.~(\ref{o6u5solv}), are known, one has at hand closed form 
expressions for related spectroscopic observables. 
Considering the $E2$ operator $T(E2) = e_B\,\Pi^{(2)}$ with $\Pi^{(2)}$ 
given in Eq.~(\ref{Pi2}), it obeys the SO(5) selection rules 
$\Delta\tau=\pm 1$ and, consequently, all $(\tau,L)$ states have 
vanishing quadrupole moments. 
The $B(E2)$ values for intraband ($g\to g$) 
transitions between states of the ground band, Eq.~(\ref{sigmaN}), 
are given by the known SO(6)-DS expressions~\cite{ibm}. For example,
\bsub
\ba
B(E2;\, g,\, \tau+1,\,L'=2\tau+2\to g,\,\tau,\,L=2\tau) &=& 
{\textstyle
e_{B}^2\,\frac{\tau+1}{2\tau+5}
(N-\tau)(N+\tau+4)} ~,
\qquad
\label{be2So6a}\\[1mm]
B(E2;\, g,\, \tau+1,\,L'=2\tau\to g,\,\tau,\,L=2\tau) &=& 
{\textstyle
e_{B}^2\,\frac{4\tau+2}{(2\tau+5)(4\tau-1)}
(N-\tau)(N+\tau+4)} ~.
\qquad\quad
\label{be2So6b}
\ea
\label{be2So6}
\esub
Similarly, the $E2$ rates for the transition connecting the pure 
spherical states, $(n_d\!=\!\tau\!=\!1,L\!=\!2)$ and 
$(n_d\!=\!\tau\!=\!0,L\!=\!0)$, satisfy 
the U(5)-DS expression of Eq. (\ref{be2nd}). 
Member states of the deformed ground band~(\ref{sigmaN}) 
span the entire $\sigma=N$ irrep of SO(6) and are not connected by 
$E2$ transitions to the spherical states since 
$\Pi^{(2)}$, as a generator of SO(6), cannot connect different 
$\sigma$-irreps of SO(6). The weak spherical $\to$ deformed 
$E2$ transitions persist also for a more general E2 operator 
obtained by adding $(d^{\dag}\tilde{d})^{(2)}$ to $T(E2)$, 
since the latter term, as a generator of U(5), cannot connect 
different $n_d$-irreps of U(5).
By similar arguments, there are 
no $E0$ transitions involving these spherical states, 
since $T(E0)$ is diagonal in $n_d$. 
These symmetry-based selection rules result in strong electromagnetic 
transitions between states in the same class, associated with a given shape, 
and weak transitions between states in different classes, hence 
can be used to identify isomeric states. 
\begin{figure}[t]
  \centering
\includegraphics[width=9cm]{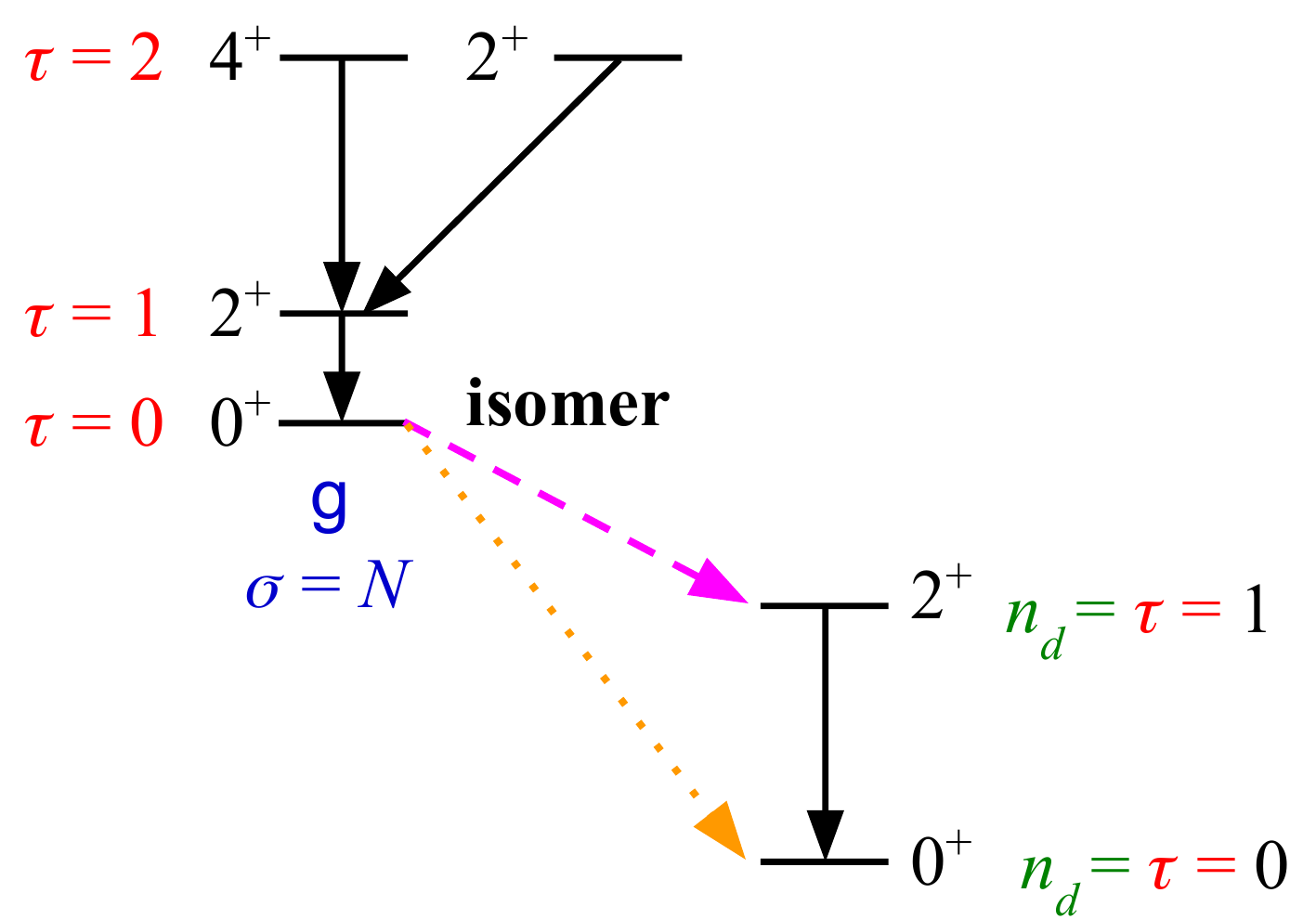}
  \caption{
Signatures of U(5) and SU(6) PDSs in S-G shape coexistence.
Strong intraband E2 transitions (solid lines) 
obey the analytic expressions of Eqs.~(\ref{be2So6}) and~(\ref{be2nd}). 
Retarded E2 (dashes lines) and E0 (dotted lines) decays 
identify isomeric states.
\label{fig14SGbe2}}
\end{figure}

The evolution of structure away from the critical point, 
where the spherical and deformed configurations are degenerate, 
can be studied by incorporating the U(5) or SO(6) Casimir operators 
in $\hat{H}'$~(\ref{HprimeSG}), still retaining the desired SO(5) symmetry. 
Adding an $\epsilon\hat{n}_d$ term, 
will leave the pure spherical $n_d=0,1$ states unchanged but will shift 
the deformed $\gamma$-unstable ground band to higher energy of 
order $\epsilon N/3$. 
Similarly, adding a small $\alpha R^{\dag}_0R_0$ term~(\ref{R0R0}), will 
leave the solvable SO(6) $\sigma=N$ ground band unchanged, but will 
shift the spherical ground state ($n_d\!=\!L\!=\!0$) to higher energy of 
order $\alpha N^2$. As discussed, the $L\!=\!0$ state of the excited 
configuration will exhibit retarded $E2$ and $E0$ decays to states of the 
lower configuration, hence will have the attributes of an isomer state, 
as depicted schematically in Fig.~14.

\section{Concluding remarks}

We have presented a comprehensive symmetry-based approach for 
describing properties of multiple shapes 
in the framework of the interacting boson model (IBM) of nuclei.
It involves the construction of a number-conserving rotational-invariant 
Hamiltonian which captures the essential features of the dynamics near the 
critical point, where two (or more) shapes coexist. 
The Hamiltonian conserves the dynamical symmetry (DS) of selected bands, 
associated with each shape. 
Since different structural phases correspond to incompatible 
(non-commuting) dynamical symmetries, the symmetries in question 
are shared by only a subset of states, and are broken in the remaining 
eigenstates of the Hamiltonian. 
The resulting structure is, therefore, that of coexisting multiple 
partial dynamical symmetries (PDSs). 

We have applied the proposed approach and examined
the relevance of the PDS notion 
to a rich variety of 
multiple quadrupole shapes, spherical and deformed (axial and non-axial). 
The shape coexistence scenarios and related PDSs considered include 
(i)~U(5)-PDS and SU(3)-PDS in relation to spherical-prolate coexistence. 
(ii)~SU(3)-PDS and $\bsu3$-PDS in relation to prolate-oblate coexistence. 
(iii)~U(5)-PDS, SU(3)-PDS and $\bsu3$-PDS in relation to triple 
spherical-prolate-oblate coexistence.
(iv)~U(5)-PDS and SO(6)-PDS in relation to coexisting spherical and 
$\gamma$-unstable deformed shapes. In each case,
the underlying potential-energy surface exhibits multiple minima which 
are near degenerate. 
As shown, the constructed Hamiltonian has the capacity to have 
distinct families of states whose properties reflect the different 
nature of the coexisting shapes. 
Selected sets of states within each family, 
retain the dynamical symmetry associated with the given shape. 
This allows one to obtain 
closed expressions for quadrupole moments and transition 
rates, which are the observables most closely related to the nuclear shape. 
The resulting analytic expressions, 
[Eqs.~(\ref{QLbe2Su3}), (\ref{QL3u5}), (\ref{quadmom}), 
(\ref{be2}), (\ref{be2nd}), (\ref{be2So6})], 
are parameter-free predictions, except for a scale, and can be used 
to compare with measured values of these observables 
and to test the underlying partial symmetries.
The purity and good quantum numbers of selected states 
enable the derivation of symmetry-based selection rules for 
electromagnetic transitions (notably, for $E2$ and $E0$ decays) and the 
subsequent identification of isomeric states. 
The evolution of structure away from the critical-point 
can be studied by adding to the Hamiltonian the Casimir operator 
of a particular DS chain, which will leave unchanged the ground band of 
one configuration but will shift the other configuration(s) 
to higher energy and may alter their symmetry properties.

The critical-point Hamiltonians obtained in the procedure of 
Eq.~(\ref{bases12}), often involve three-body interactions. 
Similar cubic terms were encountered in previous 
studies within the IBM, in conjunction with 
triaxiality~\cite{heyde84,zamfir91}, 
band anharmonicity~\cite{GarciaRamos09,Ramos00} 
and signature splitting~\cite{Leviatan13,Bonatsos88} in deformed nuclei. 
Higher-order terms show up naturally in microscopic-inspired IBM Hamiltonians
derived by a mapping from self-consistent mean-field 
calculations~\cite{nomura13,Nomura12}. 
Near shell-closure such critical-point Hamiltonian 
can be regarded as an effective number-conserving 
Hamiltonian, which simulates the excluded 
intruder-configurations by means of higher-order terms. Indeed, 
the energy surfaces of the IBM with configuration 
mixing~\cite{Frank04,Morales08,hellemans09} 
contain higher-powers of $\beta^2$ 
and $\beta^3\cos3\gamma$, as in Eq.~(\ref{surface}). 
Recalling the microscopic interpretation of the IBM bosons 
as images of identical valence-nucleon pairs, the results of the present study 
suggest that for nuclei far from shell-closure, shape coexistence can occur 
within the same valence space.

As discussed, the yrast states of each coexisting configuration, 
({\it e.g.}, the prolate and oblate ground bands) 
are unmixed and retain their individual symmetry character
({\it e.g.}, the SU(3) and $\bsu3$ character). 
This situation is different from that encountered in the 
neutron-deficient Kr~\cite{Clement07} and Hg~\cite{Bree14} isotopes, 
where the observed structures are strongly mixed. 
It is in line with the recent evidence for shape-coexistence 
in neutron-rich Sr isotopes, where 
spherical and prolate-deformed configurations exhibit very weak 
mixing~\cite{Clement16}. Band mixing can be incorporated in the present 
formalism by adding kinetic rotational terms which do 
not affect the shape of the energy 
surface~\cite{kirlev85,Leviatan87,levkir90,levshao90}.
For an Hamiltonian with one-, two- and three-body terms,
the rotational terms are of three types. (A)~Operators related to 
the Casimir operators $\hat{C}_{2}[G]$ of the groups ($G$) 
in the chain $\bso6\supset SO(5)\supset SO(3)$, 
where the generators of $\bso6$ are 
$U^{(\ell)}= (d^{\dag}\tilde{d})^{(\ell)}$, $\ell=1,3$ and 
$\bPi2=i(d^{\dag}s-s^{\dag}\tilde{d})$.
These orthogonal groups correspond to ``generalized'' rotations 
associated with the $\beta$-, $\gamma$-, and Euler angles degrees 
of freedom~\cite{Leviatan87}. 
(B)~Operators of the form $\hat{n}_d\hat{C}_{2}[G]$.
(C)~Operators of the form 
$\Pi^{(2)}\cdot (\,\bPi2\,\bPi2\,)^{(2)}$, 
$\Pi^{(2)}\cdot (U^{(1)}\,U^{(1)})^{(2)}$, 
$U^{(2)}\cdot (U^{(1)}\,U^{(1)})^{(2)}$, 
$U^{(2)}\cdot (U^{(1)}\,U^{(1)})^{(2)}$, 
$i\bPi2\cdot(U^{(2)}\,U^{(3)})^{(2)}$ and their Hermitian conjugates.
Operators in classes (A) and (B) are diagonal in the SO(5) quantum 
number $\tau$, while those in class~(C) allow for $\tau$ mixing.
Most of these rotational terms do not commute with the intrinsic part 
of the Hamiltonian hence can  shift, split and mix the bands generated 
by the latter. So far, these effects were considered only 
for the operators of class~(A) in conjunction with the coexistence 
of spherical and prolate-deformed shapes~\cite{Leviatan06}, 
hence a detailed systematic 
study of other terms is called for.
It should be noted that if the induced band-mixing is strong, it may 
destroy the PDS property of the eigenstates of the complete Hamiltonian.

Shape-coexistence in an interacting system, such as nuclei, 
occurs as a result of a competition between terms in the Hamiltonian with 
different symmetry character, which leads to considerable symmetry-breaking 
effects in most states. To address the persisting regularities in 
such circumstances, amidst a complicated environment of other states, 
one needs to enlarge the traditional concepts of exact dynamical symmetries. 
The present symmetry-based approach accomplishes that 
by employing such an extended notion of partial dynamical symmetry (PDS). 
In the same spirit that exact dynamical symmetries are known 
to serve as benchmarks for the dynamics of a single shape, it appears 
that partial dynamical symmetries have the capacity and potential 
to act as benchmarks for the study of multiple shapes in nuclei. 
PDSs can provide a convenient starting point, guidance and test-ground 
for more detailed treatments of this intriguing phenomena. 
Further work is required to quantitatively asses to what extent partial 
symmetries persist in real nuclei, where shape coexistence necessitates 
additional symmetry-breaking effects due to departures from the 
critical-point and band mixing. 
It is gratifying to note that shape-coexistence in nuclei, 
exemplifying a quantal mesoscopic system,  
constitutes a fertile ground for the development and testing 
of generalized notions of symmetry.

\ack
This work is supported by the Israel Science Foundation (Grant 586/16).\\

\end{document}